\documentclass[%
 reprint,
 amsmath,amssymb,
 aps,
]{revtex4-2}

\usepackage{graphicx}
\usepackage{dcolumn}
\usepackage{bm}
\usepackage{natbib}
\usepackage[whole]{bxcjkjatype}
\usepackage{xcolor}
\usepackage{booktabs}
\usepackage{makecell}

\usepackage{url}
\usepackage{aas_macro}

\newcommand{\syv}[1]{\textcolor{black}{#1}}
\newcommand{\TT}[1]{\textcolor{black}{#1}}

\begin{document}

\preprint{APS/123-QED}

\title{\syv{Detectability and Template Distinguishability of the Dark Ages 21 cm Global Signal with Wide and Sparse Frequency Coverage}}

\author{Shintaro Yoshiura}%
\affiliation{Institute for Advanced Research, Nagoya University, Furo-cho Chikusa-ku, Nagoya 464-8601, Japan}
 \email{yoshiura.shintaro.y5@f.mail.nagoya-u.ac.jp}
\affiliation{Graduate School of Science, Division of Particle and Astrophysical Science,
Nagoya University, Furocho, Chikusa-ku, Nagoya, Aichi 464-8602, Japan.}
\affiliation{Mizusawa VLBI Observatory, National Astronomical Observatory of Japan, 2-21-1 Osawa, Mitaka, Tokyo 181-8588, Japan}

\author{Fumiya Okamatsu}%
\affiliation{Department of Physics, College of Humanities and Sciences, Nihon University, Tokyo, 156-8550, Japan}

\author{Tomo Takahashi}%
\affiliation{Department of Physics, Saga University, Saga 840-8502, Japan}

\date{\today}

\begin{abstract}
    
The Dark Ages 21\,cm signal, observed at frequencies below 50 MHz, can serve as a powerful probe of cosmology, as the standard cosmological model predicts a well-defined 21\,cm spectral shape, \syv{while several non-standard scenarios produce distinctive spectral features.}
\syv{In this work, we present a Bayesian-evidence-based assessment of the detectability of representative Dark Ages 21\,cm signals and of the ability to discriminate among their spectral templates. We compare multiple cosmological signal templates within a common analysis framework, adopting physically motivated foreground models, optimistic error levels, and several observing strategies. This framework allows us to quantify how frequency coverage and sampling affect both the detectability and the ability to distinguish among different spectral shapes.}
\syv{Using Bayesian model comparison, we show that observations covering 1--50\,MHz provide evidence for a non-zero 21\,cm signal for all the models considered in this work. In particular, without observations below 3\,MHz, the standard cosmological signal cannot be detected even after 10,000\,h of integration, because the free--free absorption component is insufficiently constrained. Regarding template discrimination, the $\Lambda$CDM signal can be distinguished from other models except models with similar spectral shapes and an excess radio background model with a wide band observation.}
Furthermore, even with observations measured at 5 MHz intervals over the frequency range $1-50$ MHz, the 21\,cm signal can be identified if the errors are sufficiently small. This indicates that the intrinsic 21\,cm spectral shape can be captured without foreground degeneracy even with a limited number of frequency channels. \syv{These results quantify, within the idealized assumptions adopted in this work, how much of the intrinsic Dark Ages 21\,cm spectral information can be retained under limited frequency sampling. }
\end{abstract}

\maketitle


\section{\label{sec:1}Introduction}

During the Dark Ages, before the formation of the first stars, the Universe was filled with neutral hydrogen gas. The highly redshifted 21\,cm line from this gas, observed at radio frequencies, gives the only direct probe of the Dark Ages \cite{1977SvAL....3..155V, 1990MNRAS.247..510S, Furlanetto2006,2012RPPh...75h6901P}. The Dark Ages 21\,cm signal can provide a stringent test of cosmological models as it can be predicted solely from cosmology, free from astrophysical uncertainties.
In the standard $\Lambda$CDM cosmology, the signal appears as an absorption feature with a depth of $-40$\,mK at the peak frequency 15\,MHz. In \cite{2024PhRvL.133m1001O}, the consistency ratio—defined as the ratio of the signals at a few frequency points—does not change in $\Lambda$CDM even when the cosmological parameters are varied, and can effectively distinguish alternative cosmological models.

After the Dark Ages, the Cosmic Dawn and Epoch of Reionization eras follow, where the 21\,cm line can be a powerful observable to understand astrophysics such as when and what types of stars formed and how the Universe was heated. To date, several ground-based instruments have been used to probe the sky-averaged 21 cm signal;
EDGES reported a strong absorption feature at a redshift of 17.8 \cite{2018Natur.555...67B}. 
However, a follow-up observation with SARAS3 did not detect the signal \cite{2022NatAs...6..607S}, and possible systematic errors in the EDGES data have been pointed out \cite{2018Natur.564E..32H,2019ApJ...874..153B,2019ApJ...880...26S,2021MNRAS.502.4405B,2025A&A...698A.152C}. Nevertheless, the EDGES result has motivated many theoretical studies aimed at constructing models to explain the result or constraining existing models (see, e.g., \cite{2023PASJ...75S.154M} and references therein).
Interestingly, some proposed models predict an enhanced signal during the Dark Ages. Thus, the 21 cm line from this era remains a powerful probe of such models, even in light of later cosmic epochs, and follow-up observations below 50 MHz are highly anticipated.

In light of this background, several projects aim to detect the sky-averaged 21\,cm signal from the Dark Ages, using moon orbiting satellite (e.g. CosmoCube~\cite{2025RASTI...4...61A}, DAPPER~\cite{2019BAAS...51c...6B}, DARE~\cite{2017ApJ...844...33B}, DSL \citep{2021RSPTA.37990566C}, NCLE \cite{2020AdSpR..65..856B}, PRATUSH \cite{2023ExA....56..741S}, SEAMS~\cite{Tanti2023SEAMSAS}, TREED \cite{TREED, TREED_prep}) and using moon lander (e.g. DEX~\cite{2025arXiv250403418B}, FARSIDE~\cite{2021arXiv210308623B}, LARAF~\cite{2024RSPTA.38230094C}, LuSEE-Night~\cite{2023arXiv230110345B}, TSUKUYOMI~\cite{2024SPIE13092E..2LI,2024SPIE13092E..78Y}).
These projects are designed to measure the signal from outside the Earth to avoid ionospheric effects. In addition, radio-frequency interference (RFI) is also a major obstacle, which could be mitigated by observing the signal from the lunar far side.
 
Even if RFI can be reduced in this way, the most significant challenge arises from astrophysical foregrounds, dominated by synchrotron emission from our Galaxy, with contributions from free-free emission and emission from extragalactic radio sources, which can reach the noise level of $10^{5}$\,K at 15\,MHz. 
Thus, careful treatment of systematic errors is essential. For example, 
accurate instrument calibration and proper modeling of reflected-wave noise parameters are required to avoid non-smooth spectral structures \cite{2017ApJ...835...49M, 2021MNRAS.505.2638R,2022MNRAS.517.2264M,2025MNRAS.543.4312K,2026MNRAS.546ag232K}. In addition, the chromaticity of the instrumental beam can create unwanted spectral structure due to the coupling between the non-uniformity of the antenna beam, the anisotropy of the foreground emission, and the spatial coordinates of the instrument 
\citep[e.g.,][]{2014MNRAS.437.1056V,2020ApJ...905..113H,2022MNRAS.515.1580S,2023MNRAS.521.3273S}.
One of possible approaches to avoid this issue could be to measure a few frequencies with a variable-length dipole to achieve a stable instrumental beam shape over a wide frequency range, whereas a dipole measuring a wide band can suffer from large loss and beam chromaticity, which is motivated by the results of \cite{2024PhRvL.133m1001O} and is planned to be adopted in the TREED project. A similar concept for ground-based observations has also been proposed in \cite{Safari10464608}.

The first milestone for the Dark Ages 21\,cm line is the detection of a non-zero signal, regardless of accuracy or model discrimination. \TT{In this work, we first investigate the detectability of the signal. Then we further study the distinguishability of the template for example models beyond the standard $\Lambda$CDM with a physically motivated foreground modeling and assuming observing strategies motivated by recent projects. }

\syv{Previous studies have demonstrated the potential of the Dark Ages 21-cm signal both for precision cosmology within the standard $\Lambda$CDM framework \citep[e.g.][]{2010PhRvD..82b3006P, 2012RPPh...75h6901P,PhysRevD.87.043002,2023NatAs...7.1025M, 2026ApJ...997L..35B} and for probing departures from it, including excess radio backgrounds \citep[e.g.][]{2019MNRAS.486.1763F,2024MNRAS.527.1461M} and millicharged dark matter \citep[e.g.][]{2024MNRAS.527.1461M}, dark-matter--baryon interactions and other interacting dark-matter scenarios \citep[e.g.][]{2014PhRvD..90h3522T,2015PhRvD..92h3528M,2024MNRAS.527.1461M,2026MNRAS.548ag581P},  dark-matter annihilation and decay \citep[e.g.][]{2006PhRvD..74j3502F,2007MNRAS.377..245V,2018PhRvD..98b3501L,2025PDU....5002145M}, subgalactic dark-matter clumping \citep[e.g.][]{2025NatAs...9.1723P}, Early dark energy \citep[e.g.][]{2018JCAP...08..037H,
2024PhRvL.133m1001O}, superconducting cosmic strings \citep{2025PhRvD.112h3540S}, primordial black holes \citep[e.g.][]{2022PhRvD.105j3026S,2021MNRAS.508.5709Y}, and primordial magnetic fields \citep[e.g.][]{2024PhRvD.110l3506M,2026A&A...706A..48N}. }

\syv{Several of these studies translated the predicted spectral signatures into detectability forecasts and for effective integration times ranging from $10^3$ to $10^5$~h. Early Fisher-matrix studies investigated the recovery of the global signal in the presence of smooth foreground components and highlighted the severe degeneracy between the pre-stellar Dark Ages signal and foreground emission \citep[]{2010PhRvD..82b3006P,PhysRevD.87.043002}. In \citep{2024MNRAS.527.1461M}, they assessed both the detectability of excess-radio-background and millicharged-dark-matter global signals relative to a null 21-cm signal and their distinguishability from the standard signal, marginalizing over the amplitude of a single power-law synchrotron foreground component.}

\syv{Our aim is complementary \TT{to the previous studies}: rather than forecasting parameter constraints for any particular physical model, we assess how robustly representative spectral templates can be detected and distinguished across different frequency-coverage strategies and physically motivated foreground prescriptions. Throughout this work, we employ Bayesian evidence and nested-sampling techniques have been widely used in 21-cm analyses to select foreground models, assess signal-versus-null hypotheses, and diagnose instrumental systematics. e.g. \cite{2015MNRAS.449L..21H,2016MNRAS.461.2847B,2016MNRAS.455.3829H,2019MNRAS.487.1160B,2020MNRAS.492...22S,2022MNRAS.509.4679A,2022PASA...39...52S,2023PASA...40...16S, 2025MNRAS.541.2262S, 2025MNRAS.544.2340S}, as a metric for detection and template distinguishability. }

This paper is organized as follows. In  Section~\ref{sec:2}, we introduce our statistical metric and several cosmological models for the 21\,cm signal, and discuss foregrounds, instrumental noise, and observing strategies. In Section~\ref{sec:3}, we present our results and examine the detectability and template distinguishability of selected observing strategies. Section~\ref{sec:4} summarizes this work.

\section{Method}\label{sec:2}

As a statistical metric, we employ the Bayesian evidence. Here we briefly describe it by following Ref.~\cite{2020MNRAS.492...22S}. 
For a data set $D$, a model $M_i$, and model parameters $\Theta$, the posterior distribution of the parameters is given by
\begin{equation}
    P(\Theta|D, M_i) = \frac{P(D|\Theta, M_i)P(\Theta|M_i)}{P(D|M_i)} ,
\end{equation}
where $P(\Theta|M_i)$ is the prior distribution, $P(D|\Theta, M_i)$ is the likelihood, and $P(D|M_i)\equiv Z_i$ is the Bayesian evidence. To evaluate $Z_i$, we use \textsc{PolyChord} \cite{2015MNRAS.450L..61H,2015MNRAS.453.4384H}, which performs nested sampling. The Bayes factor, $ B_{i,j} = \frac{Z_i}{Z_j}$, defined by the ratio of the Bayesian evidences.  For the purpose of model comparison, the logarithmic Bayes factor of two models $ \ln B_{i,j} = \ln Z_i - \ln Z_j$, can be used. The Bayes factor quantifies the odds of model $i$ relative to model $j$ when equal prior probabilities are assigned to the models. This quantity is commonly interpreted as follows \cite{Kass:1995loi}: (1) $2 \ln B_{i,j} <2$, for which the evidence is ``not worth more than a bare mention''; (2) $2<2 \ln B_{i,j}<6$, positive; (3) $6<2 \ln B_{i,j}<10$, strong; and (4) $10<2 \ln B_{i,j}$, very strong.

The 21\,cm signal is proportional to $1 - T_\gamma/T_{\rm S}$, where $T_\gamma$ is the radio background temperature and $T_{\rm S}$ is the spin temperature. 
The spectral shape of the 21\,cm signal depends on the physical processes that affect the spin temperature and the kinetic gas temperature. The details of the 21\, signal are given, e.g., 
in \cite{Furlanetto2006,2012RPPh...75h6901P}. 

\begin{figure*}
\centering
\includegraphics[width=15cm]{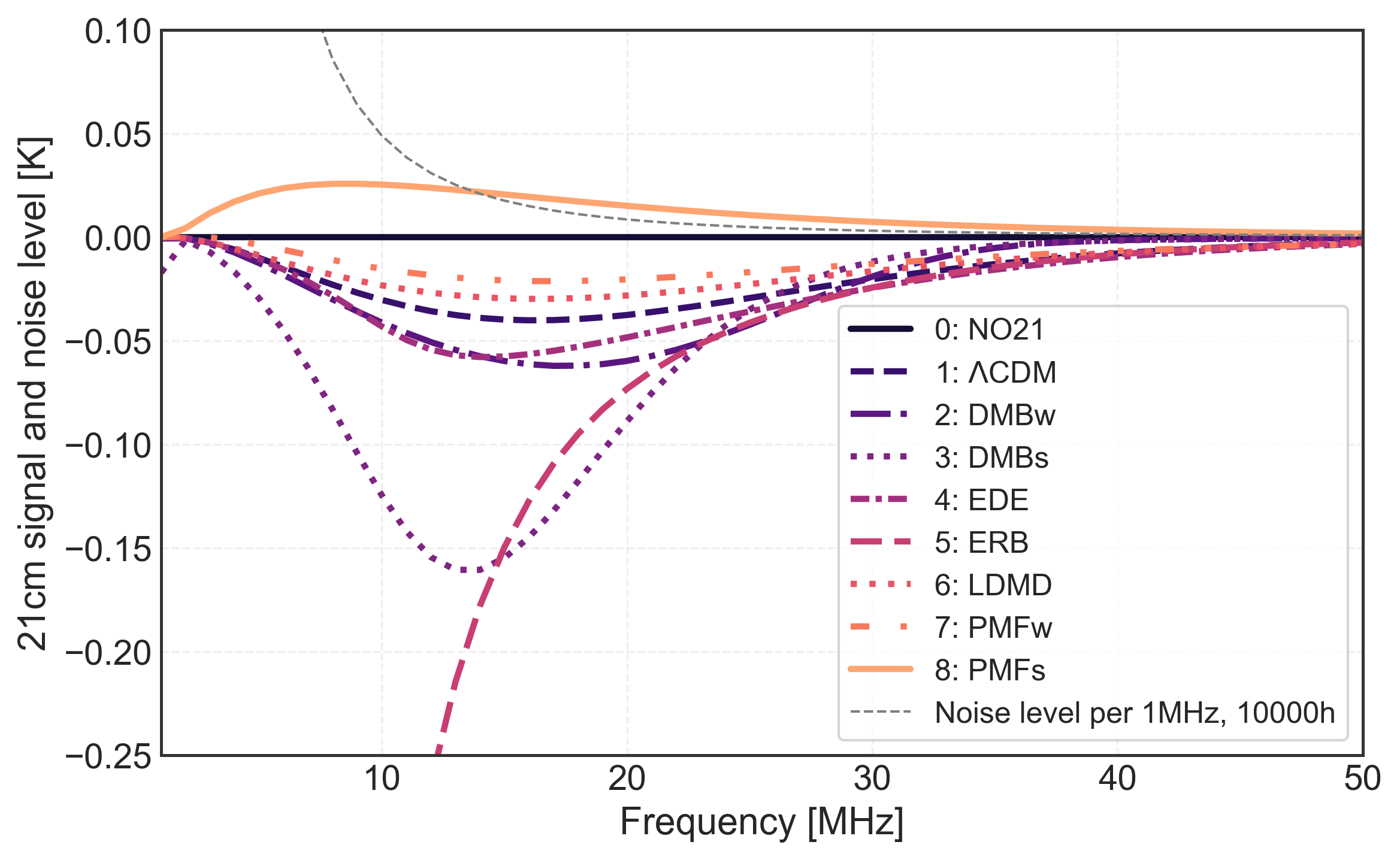} 
\caption{\label{fig:modelsplot}
The 21\,cm signal spectra used in this work for model comparison. There are nine models: eight cosmological models ($i=1$--8) and one zero-signal model ($i=0$). The thin dashed line shows the thermal-noise level assuming 10{,}000 hours of integration time.
}
\end{figure*}

In this work, we employ eight models for the 21\,cm signal in \cite{2024PhRvL.133m1001O,okamatsu_etal}.
We modified RECFAST \cite{1999ApJ...523L...1S,2000ApJS..128..407S,2008MNRAS.386.1023W,2009MNRAS.397..445S} to 
\TT{include changes to the heating and ionization, as well as possible modifications to the background evolution, in order to implement the eight different cosmological models introduced below.
} Figure~\ref{fig:modelsplot} shows these signals. In addition to the eight models labeled by $i$ ($i=1$--8), we also consider a model with no 21\,cm signal ($i=0$, NO21). For all models, the cosmological parameters are assumed as $\Omega_b h^2 =~0.02237, \Omega_mh^2 =~0.14237, H_0=~67.36\, {\rm km}\, {\rm s}^{-1}\, {\rm Mpc}^{-1}$ \cite{Planck:2018vyg} and $Y_p=~0.2436$ \cite{Hsyu:2020uqb}.

Here we briefly describe the characteristics of the 21\,cm models which will be considered in this paper. Details are given in \cite{2024PhRvL.133m1001O,okamatsu_etal}. 
\begin{itemize}
    \item \textbf{(1: $\Lambda$CDM)} This model corresponds to the standard 21\,cm signal in $\Lambda$CDM. 
    The absorption trough is centered at 16\,MHz with an amplitude of 40\,mK (e.g., \cite{2012RPPh...75h6901P,2023NatAs...7.1025M,2026ApJ...997L..35B}).
    \item \textbf{(2: DMBw)} This is the weak DM--baryon (DMB) coupling model.
    Gas cooling driven by dark matter–baryon coupling leads to a stronger absorption signal. We consider a velocity-dependent dark matter--baryon scattering cross section, $\sigma(v)=\sigma_0 v^{-4}$, where $\sigma_0$ sets the overall normalization of the interaction. Its strength is parameterized by the dimensionless quantity $\sigma_{17}$, defined as $\sigma_0=\sigma_{17} m_H \times10^{-17}\,\mathrm{cm}^2/\mathrm{g}$, with $m_H$ denoting the hydrogen mass. Here we set $\sigma_{17}=0.1$.
    \TT{For details of this model, see \cite{2014PhRvD..90h3522T}.} 
    \item \textbf{(3: DMBs)} 
    The underlying model is the same as DMBw, but with stronger cooling by assuming a larger scattering cross section between DM and baryon. We set $\sigma_{17}=1.0$.
    \item \textbf{(4: EDE)} Early dark energy (EDE) leads to earlier decoupling between photons and baryons due to the change of the expansion rate. 
    As a result, the gas temperature can be lower than in $\Lambda$CDM, and the absorption amplitude becomes deeper. We consider an EDE model for which the fractional contribution of the EDE component to the total energy density is $f_{\rm EDE}=0.4$. The onset of the decay is set by the critical redshift $z_c=150$, and the subsequent decay is described by a power-law index $p=4$.
    \TT{For details of this model, see \cite{2018JCAP...08..037H}.} 
    \item \textbf{(5: ERB)} This model assumes an excess radio background (ERB) in addition to the CMB \cite{2018ApJ...858L...9D}.
    Because the 21\,cm signal is defined as an absorption signal against the radio background, a stronger absorption signal is expected with an excess radio background. We set the relative magnitude of the additional radiation temperature to the CMB temperature to be $A_R=0.03$ \cite{2019MNRAS.486.1763F}. While higher $A_R$ is required to explain the EDGES result, we adopt lower value to keep the amplitude of 21\,cm signal consistent with other models at $\nu >$ 15\,MHz for a reasonable comparison.
    \TT{For details of this model, see \cite{2019MNRAS.486.1763F}.} 
    \item \textbf{(6: LDMD)} Light dark matter decay (LDMD) 
    heats the gas even without stars, which makes the absorption trough shallower.
    \TT{For details of this model, see \cite{2007MNRAS.377..245V}.}
    \item \textbf{(7: PMFw)} 
    Primordial magnetic field (PMF) can weakly heat the gas via decaying magnetic turbulence and ambipolar diffusion, which results in shallower absorption trough.
    We set the PMF strength to $B_0 = 0.2 ~ {\rm nG}$ and the spectral index of the PMF power spectrum to $n_B = -2.9$.
    \TT{For details of this model, see \cite{Novosyadlyj:2025zxs}.} 
    \item \textbf{(8: PMFs)} 
    The underlying model is the same as PMFw,
    however, when the heating is too effective, the 21\,cm signal appears as emission line. We set to $B_0=0.5~{\rm nG}$ and $n_B=-2.9$
\end{itemize}

{The radio background temperature is given by $T_R = T_{\rm cmb} + T_{\rm ER}\left(\nu/\nu_{78}\right)^{-2.6}$ with $T_{\rm ER}=0.03$\,K for the ERB model, and by $T_R = T_{\rm cmb}$ for the other models.} For more details of model and parameter definitions, please refer to the references cited in the description of each model.

This work does not aim to interpret the physical origin of the 21 cm models, but instead examines the signal shape to investigate which types of 21 cm \TT{templates} can be detected or differentiated.
Table~\ref{tab:model} summarizes the peak amplitude and the peak frequency, except for ERB, because the ERB model does not exhibit an absorption trough. Most models have their peak at around 15\,MHz. Compared with $\Lambda$CDM, models such as DMBw, DMBs and EDE exhibit strong absorption, whereas LDMD and PMFw show weak absorption. An exception is the PMFs model which produces an emission feature at around 15 MHz.

\begin{table}[t]
\centering
\caption{Summary of spectral features (the absorption peak and the peak frequency) of the 21\,cm models employed in this work. \label{tab:model}}
\begin{tabular}{lrrr}
\toprule
Model &
\makecell{ peak [mK]} &
\makecell{ peak $\nu$ [MHz]} &
\makecell{ } \\
\midrule
1: $\Lambda$CDM & -40.1 & 16 &  \\
2: DMBw & -62.1 & 17 &   \\
3: DMBs & -160.6 & 14 &  \\
4: EDE & -57.9 & 14 &   \\
5: ERBs & - & - &   \\
6: LDMD & -29.8 & 16 &  \\
7: PMFw & -21.3 & 17 &   \\
8: PMFs & 25.7 & 9 &   \\
\bottomrule
\end{tabular}
\end{table}

For the likelihood analysis, we fix the \syv{shape of} model and cosmological parameters for each 21\,cm model and evaluate the detection and \syv{template} distinguishability performance, which is the primary objective of this work. \syv{We employ a normalization parameter $A$ to vary overall amplitude of the 21\,cm signal. } We note that fixing the cosmological parameters does not affect our conclusions, since the shapes of the 21\,cm signals in $\Lambda$CDM and in models beyond $\Lambda$CDM vary only weakly within the parameter ranges allowed by current observations \cite{2024PhRvL.133m1001O}. 

In principle, the nested-sampling algorithm can be used to infer posterior distributions of the model parameters for each extended 21\,cm scenario. However, accurate predictions for some models are computationally expensive, making a full nested-sampling analysis impractical. Therefore, in this work, we instead test the feasibility of \syv{template distinguishability} \syv{by varying overall amplitude} for a representative set of 21\,cm signal models. Recently, \citet{2024PhRvD.109l3541G} suggested a significant speed-up of nested sampling using neural ratio estimation. GPU-accelerated nested sampling can also help speed up the calculations \citep{2026arXiv260123252Y}. Implementing these new methods is beyond the scope of this work, and we leave this for future work.

In addition to the 21\,cm signal, the foreground should also be included as it dominates the signal. Although polynomial functions are also commonly used to model the foreground spectrum, they may carry some risk of producing non-physical spectra. In practice, \syv{template distinguishability} can be performed with multiple foreground models \citep{2015MNRAS.449L..21H,2020MNRAS.492...22S}. In this work, we adopt a physically motivated foreground model based on \cite{1979MNRAS.189..465C}:
\begin{eqnarray}
    T_{\rm FG} = &T_{\rm G}& (\nu/\nu_{10})^{-\alpha-2} \left(\frac{1-\exp{(-\tau_{\nu})}}{\tau_{\nu}} \right) \notag \\
    && \qquad + ~ T_{\rm eg} (\nu/\nu_{10})^{-\beta-2} \exp{(-\tau_{\nu})},
    \label{eq:fg}
\end{eqnarray}
$\nu_{10}=10\,\rm MHz$ and $\tau_{\nu}=F\,\nu^{-2.1}$. 

The first term corresponds to synchrotron emission from the Galaxy, while the second term describes the extragalactic foreground component subject to free–free absorption. We adopt an optical depth of $\tau_{\nu}=F\,\nu^{-2.1}$. The optically thick region also attenuates the 21 cm signal and the radio background \cite{2022MNRAS.513.5125S}.

\begin{figure}
\centering
\includegraphics[width=8cm]{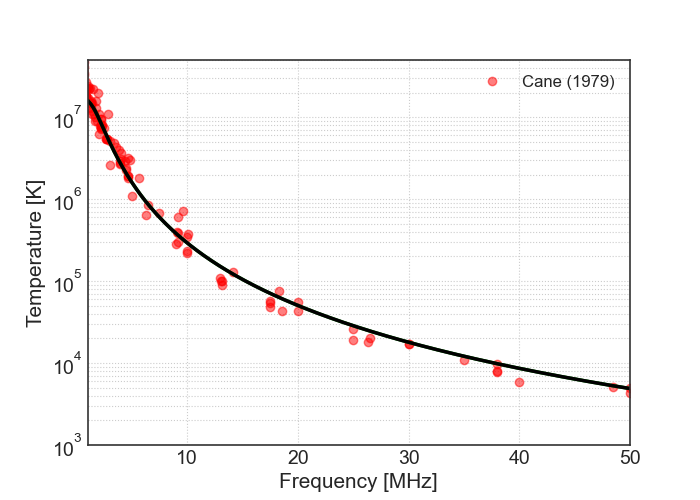}
\caption{
The solid line shows the foreground model used to generate the mock data. The red dots show the observed values from \cite{1979MNRAS.189..465C}.
}
\label{fig:foregroundmodel}
\end{figure}

As the mock foreground temperature, which is used as true foreground model $f_{\rm t}$,  the following parameter values are assumed \cite{1979MNRAS.189..465C}: ($T_{\rm G}$, $T_{\rm eg}$, $\alpha$, $F$, $\beta$ $)= (2.45\times 10^5\,\rm K$, $5.5 \times 10^4\,\rm K$, 0.515, 5, 0.8).
Figure~\ref{fig:foregroundmodel} shows the mock foreground model and the observed values listed in \cite{1979MNRAS.189..465C}. 
It should be noted that 
the foreground temperature reaches $10^7$\,K at 1\,MHz.

Designing instruments optimized for observing the Dark Ages is challenging because of the large frequency dynamic range. Thus, non-optimized instruments may have low radiation efficiency. To reduce systematics, it can be useful to introduce an attenuator between an antenna and a receiver \cite{2017ApJ...835...49M,2024MNRAS.530.4125M} while the attenuator adds noise. Here, we allow the receiver noise temperature, including attenuator noise, to be as high as 600\,K, which is motivated by \cite{2025NatSR..1534335L}.

Combining the foreground and instrumental noise, the thermal-noise level is then given by $\sigma = (T_{\rm sky} + T_{\rm rec}) / \sqrt{\Delta t\,\Delta \nu}$,
where the sky temperature is $T_{\rm sky}=f_{\rm t}$, the receiver noise is $T_{\rm rec}$, $\Delta t$ is the integration time per channel, and the channel width is $\Delta \nu = 1\,\rm MHz$.

It is not easy to predict and model instrumental systematic errors, such as calibration errors \cite{2017ApJ...835...49M}, ionospheric effects \cite{2014MNRAS.437.1056V}, and beam chromaticity \cite{2021MNRAS.506.2041A}. In this work, to investigate the ideal detectability, we do not include complicated systematic errors.  Thus, the noise level in a channel can be as low as 1.7 mK even when the sky temperature is 10,000 K.
\TT{However, systematic errors would set a minimum noise level, and thus}
in Section~\ref{sec:3}, we discuss the effect of imposing a minimum noise level on the detection of the 21 cm signal.

For the observing configurations, we assume the following three cases:
(1) observations from 1\,MHz to 50\,MHz with a wide-band antenna and a channel width of 1\,MHz;
(2) observations from 1\,MHz to 46\,MHz at 5\,MHz intervals, using an optimized antenna for each frequency channel with a bandwidth of 1\,MHz;
(3) observations from \syv{minimum frequency, $f_{\rm min}$,} to 50\,MHz with a wide-band antenna and a channel width of 1\,MHz. For configurations~(1) and (3), a wide-band antenna may introduce unwanted spectral structure due to instrumental chromaticity. For configuration~(2), we assume independent instruments optimized for each channel, which may reduce systematic errors. These strategies are motivated by the findings of \cite{2024PhRvL.133m1001O}.

Using models of the 21 cm signal, foregrounds, and noise, together with the assumed observing configurations, we assess the detection performance and \syv{template distinguishability} via Bayesian evidence. For this purpose, we construct mock data based on the 21\,cm signal of model $i$, as described in the the previous section. The mock signal at the $k$-th frequency channel,  including foreground, is given by
\begin{equation}
    D_{i,k} = (m_{i} + T_{R,t})\exp(-\tau_{\nu}) + T_{\rm FG}(\hat{\Theta}) + \syv{N_k}\,,
\end{equation}
where $m_{i}$ is the true 21\,cm signal for model $i$, $T_{R,t}$  denotes radio background temperature and $\hat{\Theta}$ is fiducial parameter of foregrounds. \syv{We added Gaussian noise $N_k$ with zero mean and standard deviation.} \syv{We note that, in practice, the effect of free--free absorption depends on the redshift of the source signal \TT{(for the 21\,cm signal \citep[e.g.,][]{2018PhRvD..98d3520B}).} In this work, however, we assume that the observed spectrum is dominated by Galactic foreground components.} We then fit the model $j$ to the mock data constructed from model $i$ \syv{with amplitude parameter $A$.} The model prediction at the $k$-th frequency channel is given by
\begin{equation}
    M_{j, k} = (A\,m_j + T_{R,j})\exp(-\tau_{\nu}) + T_{\rm FG}(\Theta) \,,
\end{equation}
where $m_{j}$ denotes the 21 cm signal, $T_{R,j}$ represents the radio background temperature, and ${\Theta}$ denotes the foreground parameters, with uniform prior $1.95\times 10^5 < T_{\rm G} < 2.95\times 10^5$,
$5.0\times 10^4 < T_{\rm eg} < 6.0\times 10^4$,
$0.4<\alpha<0.6$,
$4<F<6$,
and $0.5<\beta<1.0$. We note that, except for the EBR model, $T_{R,t}$ and $T_{R,s}$ are set equal to $T_{\rm cmb}$. 

\syv{Here, we vary only the signal amplitude while keeping the spectral shape fixed, because our aim is to investigate template-based detectability and model comparison rather than detailed signal reconstruction. We adopt a uniform prior, $0 \leq A \leq 2$. As discussed in previous studies, variations in the standard cosmological parameters primarily affect the overall normalization of the global 21\,cm signal, while producing only minor changes in its spectral shape. Under current Planck constraints, the corresponding uncertainty in the global-signal amplitude is expected to be less than approximately 15\% at the $5\sigma$ level \citep{2024PhRvL.133m1001O}. Nevertheless, we adopt the much broader upper bound of $A=2$ as a conservative prior that encompasses variations well beyond those expected within the standard cosmological model. The lower bound includes the null-signal case, $A=0$.}

\syv{The detection significance could also be assessed from the posterior distribution of $A$. In this work, however, we focus on Bayesian model comparison using the evidence, or equivalently the Bayes factor. Accordingly, for the $j=0$ model, which assumes that no global 21\,cm signal is present, we do not introduce the amplitude parameter $A$. We compare its Bayesian evidence directly with that of models that include a global-signal component. This formulation naturally incorporates the Occam penalty: models containing the additional free parameter $A$ are favored only when the resulting improvement in the fit is sufficient to compensate for the larger prior volume.}

Then we define a Gaussian likelihood as
\begin{eqnarray}
\ln L_{j}(\Theta)
= 
-\frac{1}{2}\sum_k \left[
\ln\!\left(2\pi\sigma_k^2\right)
+ \frac{\left(D_{k}-M_{k}(\Theta)\right)^2}{\sigma_k^2}
\right] \,,
\end{eqnarray}
with $\sigma_k$ is the noise level at the $k$-th frequency channel. With this likelihood, we assess the detectability and model discriminability in the next section. \syv{We note that, for each pair of models $i$ and $j$, we repeat the nested-sampling analysis for 10 independent noise realizations and report the mean $2\ln B_{i,j}$ over these realizations.}

\section{Results \& Discussion}\label{sec:3}

\subsection{Detectability}

In this section, we discuss the detectability for several observing strategies described in the previous section. Specifically, using mock data generated from the true 21\,cm model $i$, we compare the evidence $\ln Z_i$, obtained with fitted model $j=i$, with $\ln Z_0$ for the NO21 model $(j=0)$. If 2$ \ln B_{i,0} = 2(\ln Z_i - \ln Z_0$) is greater than 2, we interpret the observation as detectable, since the data favor a model $i$ suggests that a model with a non-zero signal is more supported by the data than the model without a 21\,cm signal. 

This ``detectability" should be interpreted as evidence-based preference rather than a frequentist detection significance.

Table~\ref{tab:cond1_logz} lists the results for observations covering 1\,MHz to 50\,MHz continuously with a wide-band antenna with assumption of integration times of \TT{100\,h, 1,000\,h, 10,000\,h and 100,000\,h}. \TT{The results show that the observing configurations with integration time of 1,000\,h or longer have strong detectability for most 21\,cm template models}, 
except for the PMFs model. An integration time of 1,000 hours is more realistic than the assumed 10,000 hours. Although the significance is reduced for 1,000 hours, the observation can still strongly detect all 21 cm models except PMFs. \syv{With unrealistic integration time (100,000h), the PMFs model can \TT{also} be detected.}

\begin{table}[t]
\centering
\caption{Summary of $2\ln B_{i,0}$ for observations covering 1\,MHz to 50\,MHz with an integration time of 1{,}000 hours, 10{,}000 hours and 100{,}000 hours.}
\label{tab:cond1_logz}
\begin{tabular}{lrrrr}
\toprule
Model &
\makecell{100 h} &
\makecell{1,000 h} &
\makecell{10,000 h} &
\makecell{100,000 h} \\
\midrule
1: $\Lambda$CDM & 1.9 & 10.2  & 77.5 & 722.9 \\
2: DMBw & 4.3 & 24.3 & 221.6 & 2137.7 \\
3: DMBs & 2.8 & 18.0  & 177.3 & 1760.7 \\
4: EDE  & 1.7 & 11.7  & 97.5 & 907.1 \\
5: ERB  & 0.3 & 5.0  & 30.6 & 268.7 \\
6: LDMD & 0.6 & 5.1  & 31.5 & 291.0 \\
7: PMFw & 1.1 & 2.4  & 13.8 & 119.9 \\
8: PMFs & -0.1 & -0.2  & 0.0 & 16.5 \\
\bottomrule
\end{tabular}
\end{table}

\begin{figure*}
\centering
\includegraphics[width=8cm]{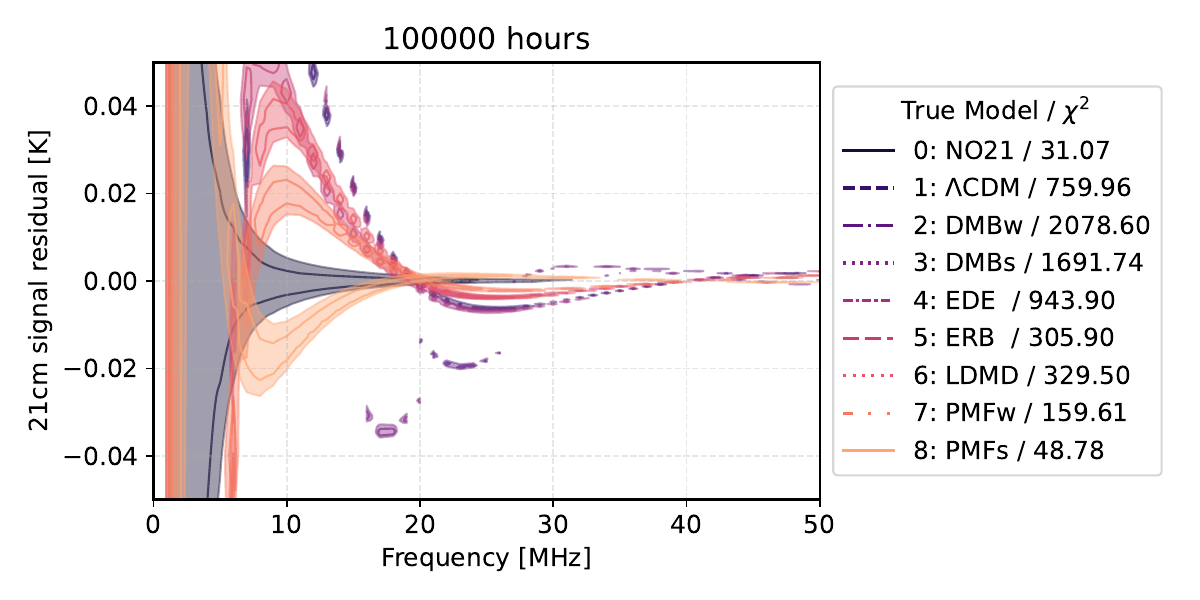}
\includegraphics[width=8cm]{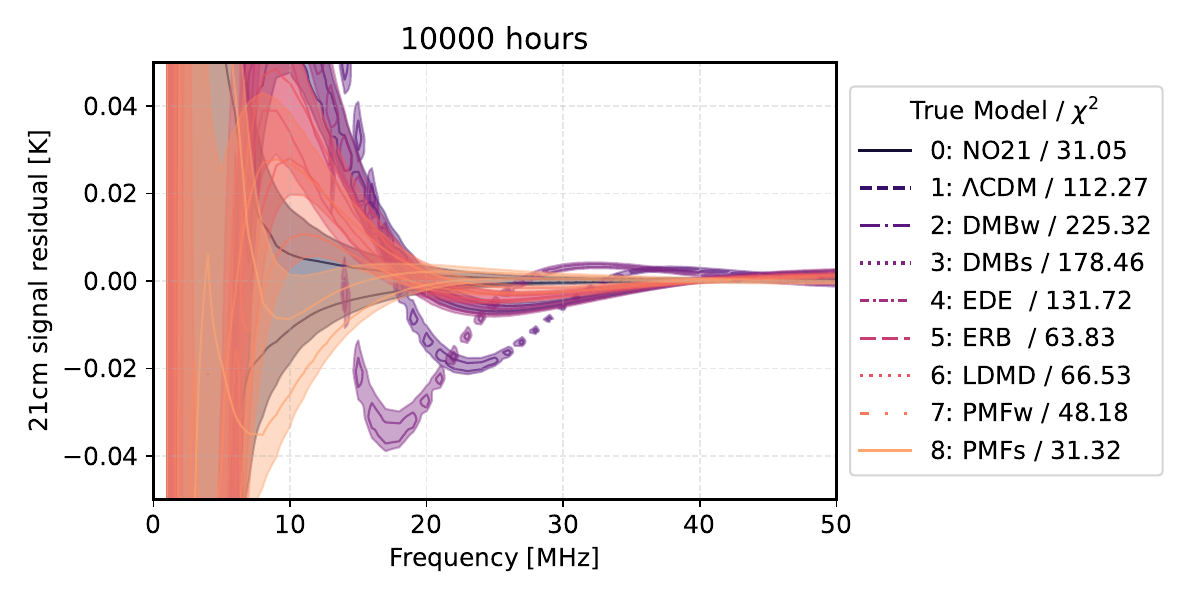}
\includegraphics[width=8cm]{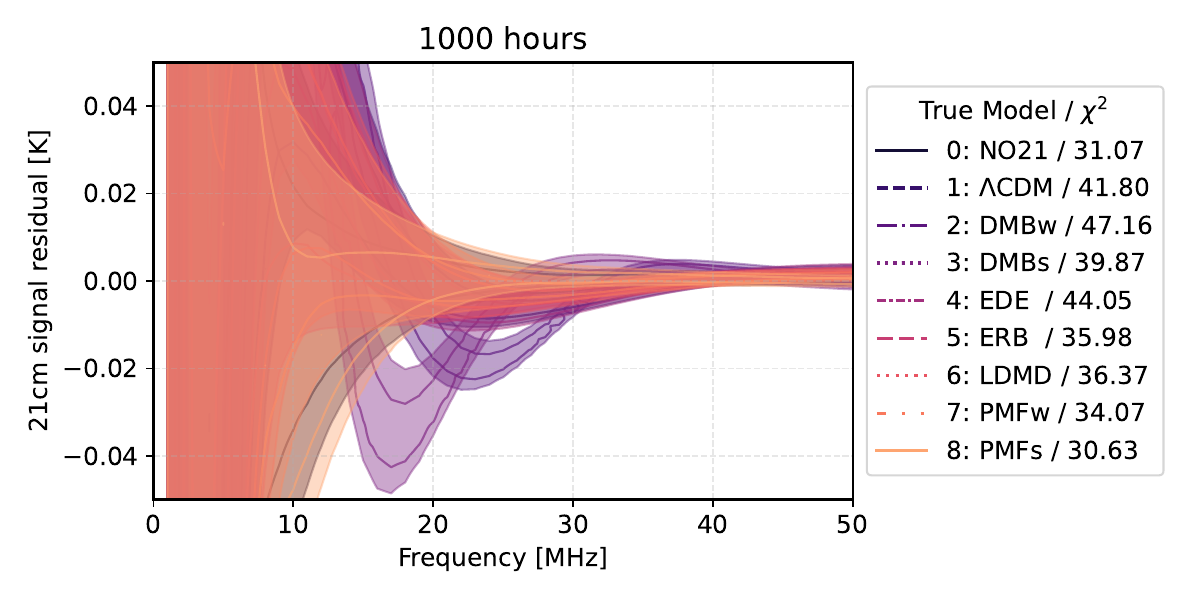}
\includegraphics[width=8cm]{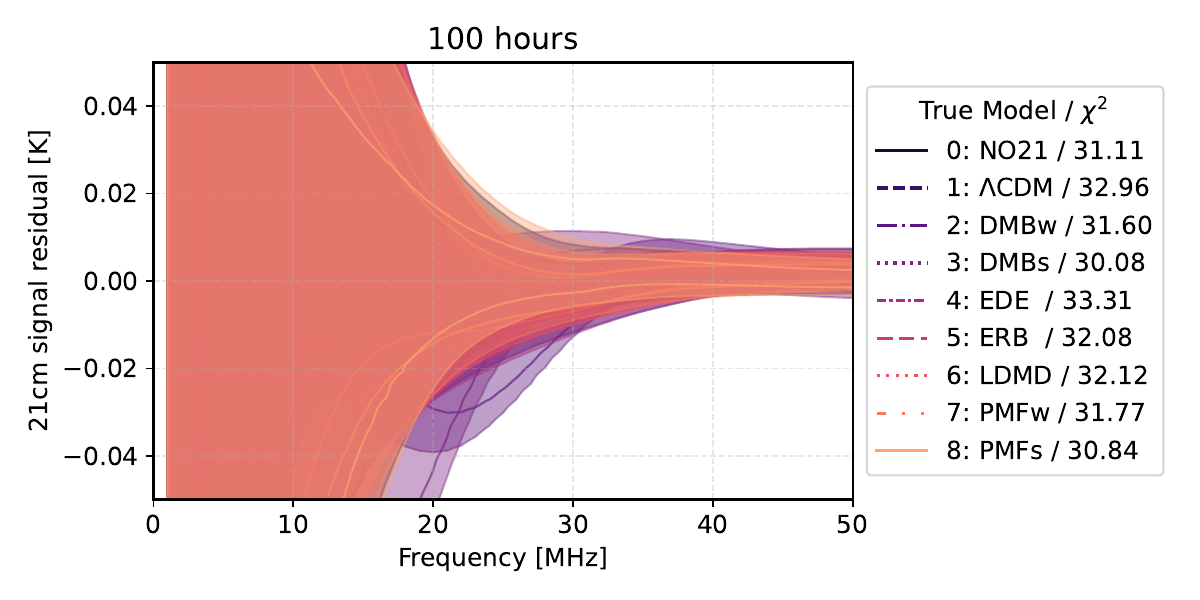}
\caption{
Residuals after foreground subtraction using the maximum-likelihood parameter sample for the case $j=0$. The contour shows 1 $\sigma$ error which is plotted using fgivenx \citep{fgivenx}. The assumed integration times are 10,000 h, 1,000 h and 100 h from \TT{top left to bottom right.}
}
\label{fig:result1maxlike}
\end{figure*}

To interpret the results more carefully, Figure~\ref{fig:result1maxlike} shows \syv{the posterior} of residual signal obtained by subtracting the foreground, evaluated at the maximum-likelihood parameters, using the NO21 as subtracted model data ($j=0$). \syv{For plotting the posterior residual, we used all samples from the results of 10 noise realizations. As we subtract zero 21\,cm signal ($j=0$), the residual indicates how much the foreground model can absorb the shape of true 21\,cm signal.} \syv{For all models, with 100,000\,h observation,} the residual signal after foreground subtraction exhibits a significant deviation from zero in the 5–30 MHz range. \syv{With 10,000 hours of observation, in the PMFs model, however, the contour is statistically consistent with that of the NO21 model, which reflects the noise level. Because the PMFs model has a weak, positive, and smooth spectrum at $\nu~>$10 MHz, the signal can be easily absorbed by the foreground model. With shorter observation times, the contour becomes broader, making it difficult to detect the 21\,cm signal even in the presence of strong absorption.}

We next assume observations at 5\,MHz intervals spanning from 1\,MHz to 46\,MHz. This observing strategy is motivated by \cite{2024PhRvL.133m1001O}, which demonstrates \TT{whether} a few frequency points are sufficient to distinguish \TT{the templates} 
between cosmological models. This observing strategy corresponds either to measurements at a small number of targeted frequencies using independently optimized instruments, or to wide-band observations in which many channels are lost due to RFI. The results for this strategy are given in Table~\ref{tab:cond2_logz}, showing that the detectability is lower than that in Table~\ref{tab:cond1_logz}. Nevertheless, with 10,000\,h integration \TT{time}, the $\Lambda$CDM, DMBw, DMBs,  EDE and LDMD models can be detected with this observing configuration, indicating that observations at a few frequency channels are in principle sufficient to detect the 21 cm signal. 

\begin{table}[t]
\centering
\caption{Same as Table~\ref{tab:cond1_logz}, but assuming measurements at 1, 6, 11, 16, 21, 26, 31, 36, 41, and 46\,MHz.}
\label{tab:cond2_logz}
\begin{tabular}{lrrrr}
\toprule
Model &
\makecell{100 h} &
\makecell{1,000 h} &
\makecell{10,000 h} &
\makecell{100,000 h} \\
\midrule
1: $\Lambda$CDM & -0.4 & -0.2  & 2.6 & 25.2 \\
2: DMBw & 0.0 & 0.5 & 7.7 & 92.0 \\
3: DMBs & 0.2 & 1.2  & 15.6 & 158.5 \\
4: EDE & 0.3  & 1.2  & 2.3 & 24.3 \\
5: ERB & 0.1 & 0.1  & 1.7 & 6.7 \\
6: LDMD & -0.2 & 0.3  & 2.0 & 11.7 \\
7: PMFw & -0.4 & 0.2  & 1.5 & 5.4 \\
8: PMFs & -0.8 & 0.2  & 0.0 & 0.1 \\
\bottomrule
\end{tabular}
\end{table}

A closer examination of ERB and PMFw in Table~\ref{tab:cond2_logz} with 10,000 h observation where $2\ln B_{i,0} > 1$, suggesting that improving the sensitivity enhances detectability. For demonstration purposes, we assume an integration time of 100,000 hours for measurements at 5 MHz intervals, in order to compare with the wide-band observation although, in practice, such a long observation is not feasible without tens to hundreds of antennae. We find that $2\ln B_{i,0}$ is greater than 5 for all models except PMFs (see Table.~\ref{tab:cond2_logz}).
This result indicates that the degradation in detectability is primarily due to sensitivity. \syv{Under the foreground and noise assumptions adopted in this study, our results demonstrate that sufficiently sensitive observations at only a few limited frequency channels can robustly detect the global 21 cm signal despite the presence of bright foreground emission.}

\syv{We find that the 21\,cm signal can be identified from a limited number of frequency samples, provided that the measurement sensitivity is sufficiently high. However, as shown in Figure~\ref{fig:result1maxlike}, significant residuals remain at the lowest frequencies when this part of the spectrum is not adequately sampled. Extending the frequency coverage to lower frequencies is therefore essential. More generally, global-signal experiments require sufficiently broad frequency coverage to capture the characteristic spectral features of the signal and reduce degeneracies with foreground emission \citep{PhysRevD.87.043002}. Because the Dark Ages 21\,cm signal typically reaches its maximum absorption amplitude at around 15\,MHz, observations must cover this spectral feature in order to constrain the signal robustly. Therefore,} for detecting the Dark Ages 21\,cm signal, it is a common strategy to observe from outside the Earth to avoid ionospheric reflection. Nonetheless, even though ionospheric refraction and RFI cannot be fully avoided, ground-based observations can probe frequencies down to 10–20 MHz \citep[e.g.][]{2024arXiv240109096K}. Thus, we also assess the detectability of a wide-band observation \syv{by changing the minimum frequency range from 1 to 20 MHz}.

\begin{figure}
\centering
\includegraphics[width=8cm]{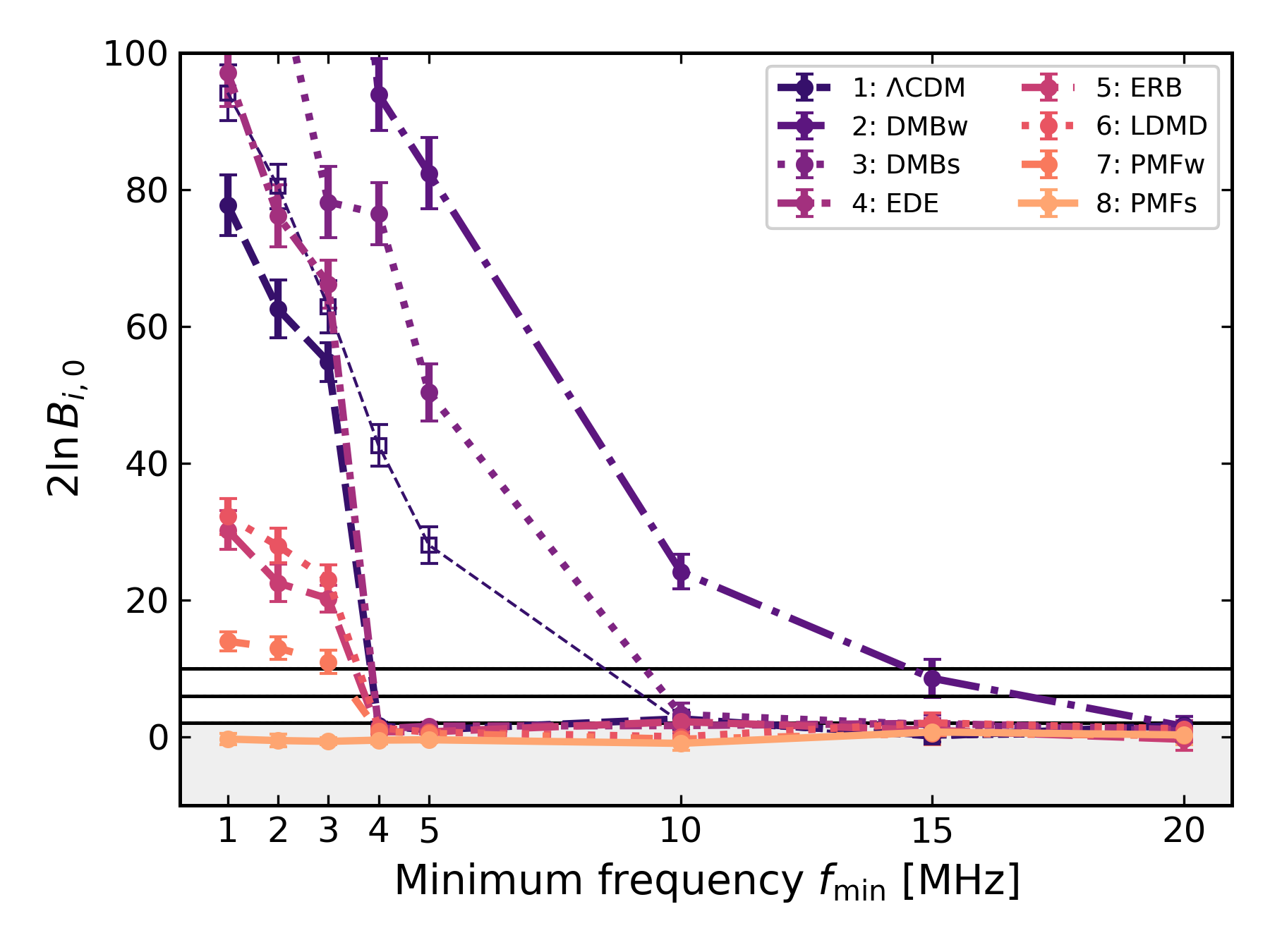}
\caption{\syv{Dependence of $2\ln B_{i,0}$ on the assumed minimum observing frequency, varied from 1 to 20\,MHz. We assume 10,000\,h of wide-band observations. The dashed curve shows the result for the $\Lambda$CDM model when free--free absorption is fixed ($F=5$). The three horizontal dotted lines indicate $2\ln B_{i,0}=2$, 6, and 10. The error bars represent the standard error estimated from 10 independent noise realizations.
}
}
\label{fig:detection_minimumfreq}
\end{figure}

\syv{Figure~\ref{fig:detection_minimumfreq}} shows the results. Notably, this observing configuration fails to detect any of the 21\,cm models except DMBw and DMBs \syv{with minimum frequency larger than 3 MHz}. \syv{This suggests that, under our assumptions, wide-band observations extending to frequencies at which free--free absorption becomes significant are essential for distinguishing the 21\,cm signal from the foreground. This result further motivates space-based measurements at such low frequencies to probe the Dark Ages signal.} 

\begin{figure}
\centering
\includegraphics[width=8cm]{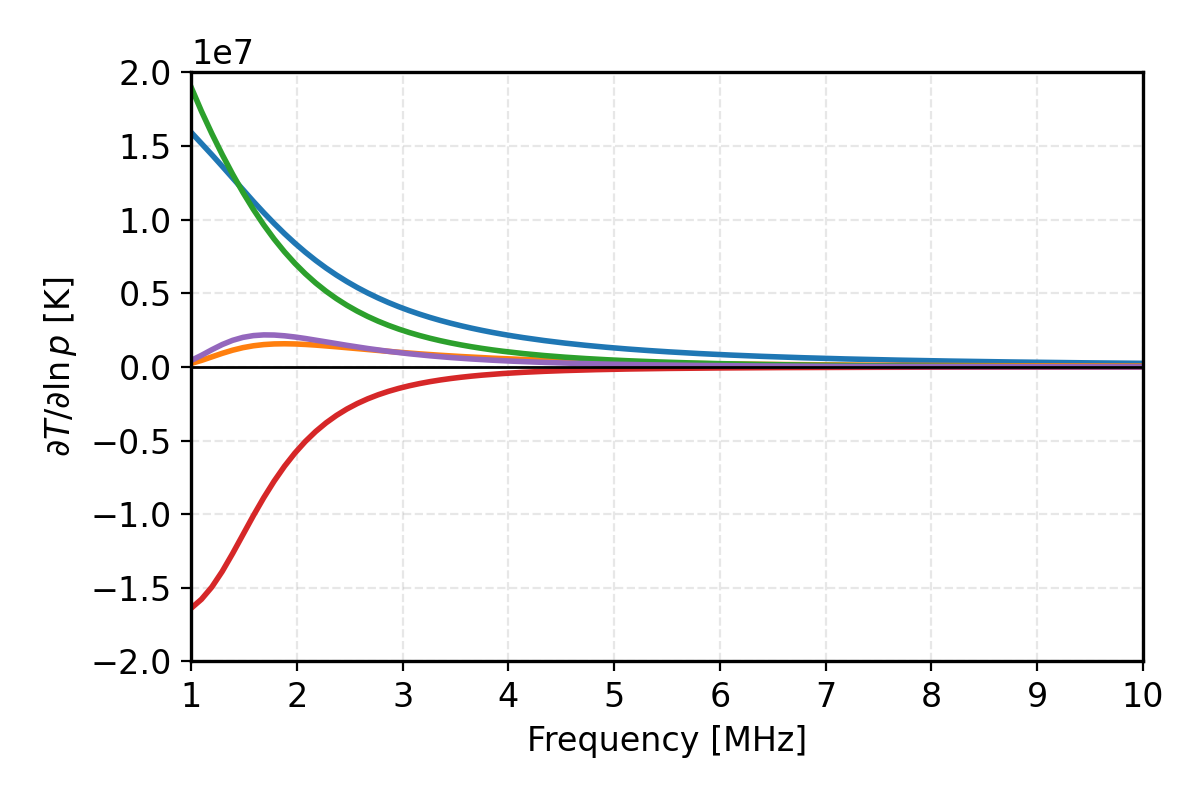}
\includegraphics[width=8cm]{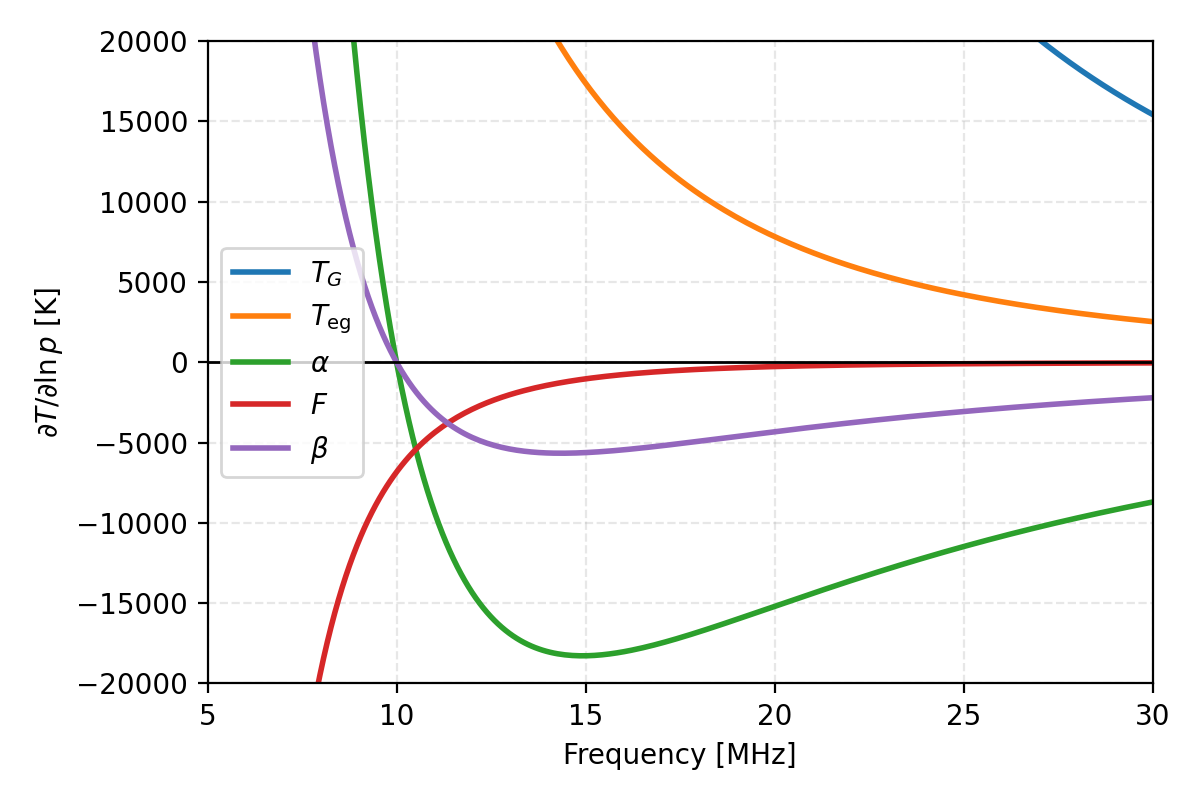}
\caption{
\syv{The logarithmic derivatives of foreground model of each foreground parameter.}
}
\label{fig:detection_minimumfreq_dev}
\end{figure}

\syv{For comparison, we test the detectability of $\Lambda$CDM model with fixed $F=5$ and omit the parameter from our analysis. 
Then, even with minimum frequency of 5 MHz, the observation can detect global signal. 
\TT{This implies that}
the free-free absorption introduces additional spectral degrees of freedom that can become degenerate with the broad curvature of the 21\,cm signal. In Figure~\ref{fig:detection_minimumfreq_dev}, we show the logarithmic parameter responses of the foreground model, defined as $\partial T_{\rm fg}/\partial \ln p_i$. Interestingly, the $F$ parameter does not introduce a spectral feature similar to the global 21,cm signal, but rather smoothly suppresses the foreground emission toward lower frequencies. In contrast, variations in $\alpha$ and $\beta$ can generate broad negative spectral features that resemble the 21\,cm absorption signal, although their amplitudes continue to increase toward lower frequencies. This indicates that the degeneracy with the global 21\,cm signal arises not from a single foreground parameter, but from the combination of multiple parameters that collectively introduce similar spectral curvature
\TT{although} we use relatively simple parametrization of foreground. In practice the strength of free-free absorption depends on direction \citep{2022MNRAS.513.5125S, 2025ApJ...994..226A}. This analysis essentially motivate the importance of measuring and modeling of the free-free absorption \citep{2021ApJ...914..128C,2022ApJ...940..180C,2022ApJ...929...32S}. } 

\begin{figure}
\centering
\includegraphics[width=8cm]{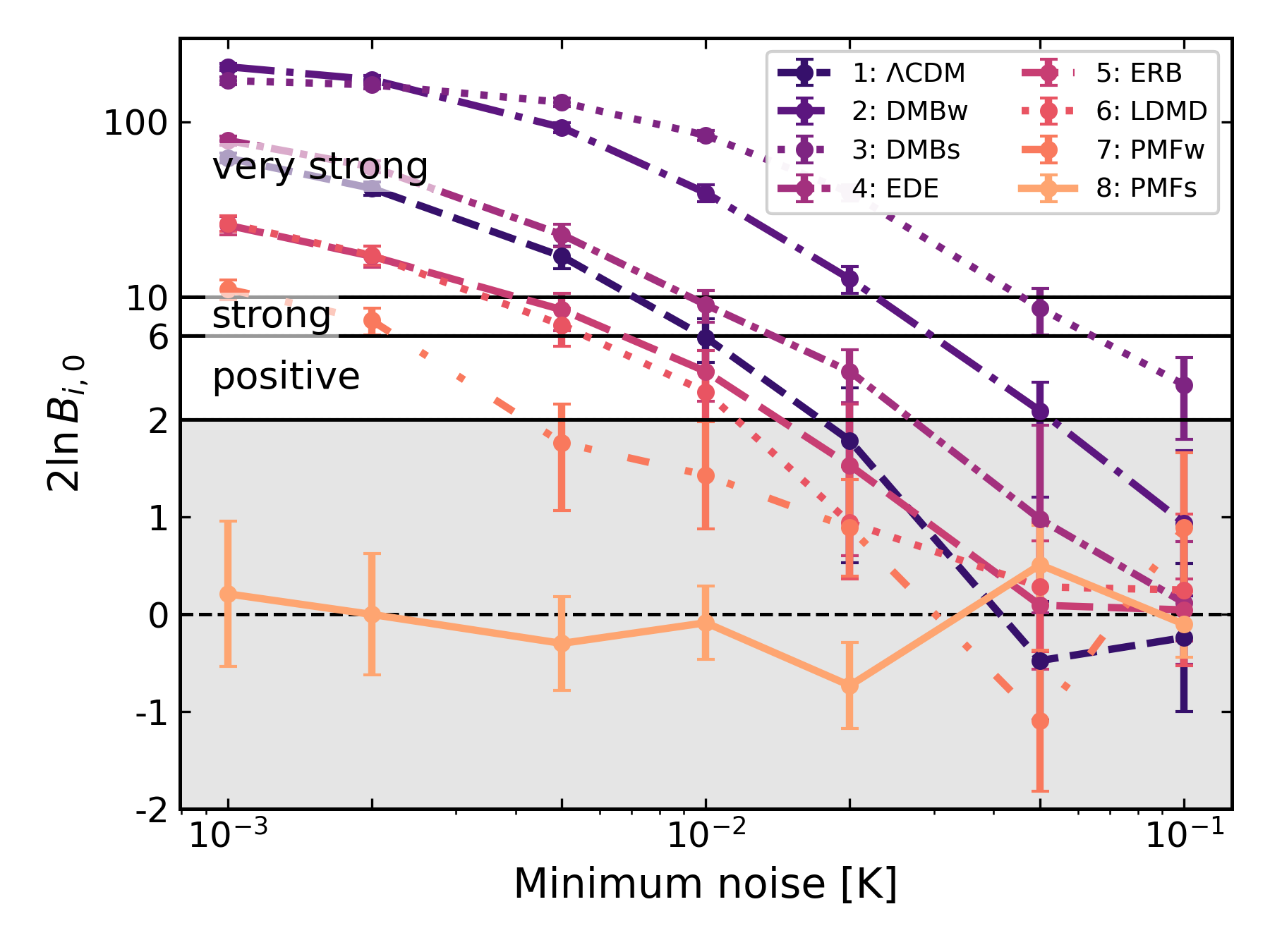}
\caption{
Dependence of $2\ln B_{i,0}$ on the assumed minimum noise level in each 1\,MHz channel. The three horizontal lines indicate the interpretation thresholds for $2\ln B_{i,0}$.
}
\label{fig:detection_minimumnoise}
\end{figure}

We next assess the minimum noise level required to detect the 21\,cm signal. So far, we have assumed only thermal noise, which can be reduced by increasing the integration time. In practice, however, systematic errors set a minimum noise level, thereby reducing the detectability of the signal. \syv{To assess the impact of a minimum noise level, we replace the thermal-noise standard deviation with the quadrature sum of the thermal noise and a frequency-independent minimum noise floor, $\sigma_k=\sqrt{\sigma_{{\rm th},k}^2+\sigma_{\rm min}^2}.$} Figure~\ref{fig:detection_minimumnoise} shows $2\ln B_{i,0}$ as a function of the minimum noise levels, for a wide-band observation from 1\,MHz to 50\,MHz. 
With a noise level of 100\,mK, none of the models are detected with strong evidence ($2\ln B_{i,0}>6$). The PMFs model remains undetectable even at a 1\,mK noise floor. 
\TT{On the other hand,}
the $\Lambda$CDM model can be detected with strong evidence ($2\ln B_{i,0}>5$) \TT{even} at 10 mK. 

\syv{Because real instrumental systematics are generally frequency dependent, our assumption of a frequency-independent noise floor is likely optimistic. Nevertheless, this result suggests that the overall level of residual systematic errors must be controlled to approximately 10 mK or below in order to detect the standard Dark Ages signal with high confidence. Modeling realistic systematic effects, such as beam chromaticity, receiver calibration errors, impedance mismatch, or sinusoidal bandpass ripples, requires adopting a specific antenna and receiver design and is therefore beyond the scope of this work. We leave a quantitative investigation of these effects within our framework to future work.}

\subsection{\syv{Template Model} Discriminability}

In the previous subsection, we focused on $2\ln B_{i,0}$ for the mock $i=1$--$8$, using the fitted data constructed from no 21\,cm signal model $(j=0;~ {\rm NO21})$, and discussed the detectability. In this section, we present results on the \syv{template model} discriminability by generating mock data from each model $i$ and fitting a different model $j~ (\ne i)$.

\begin{figure*}
\centering
\includegraphics[width=15cm]{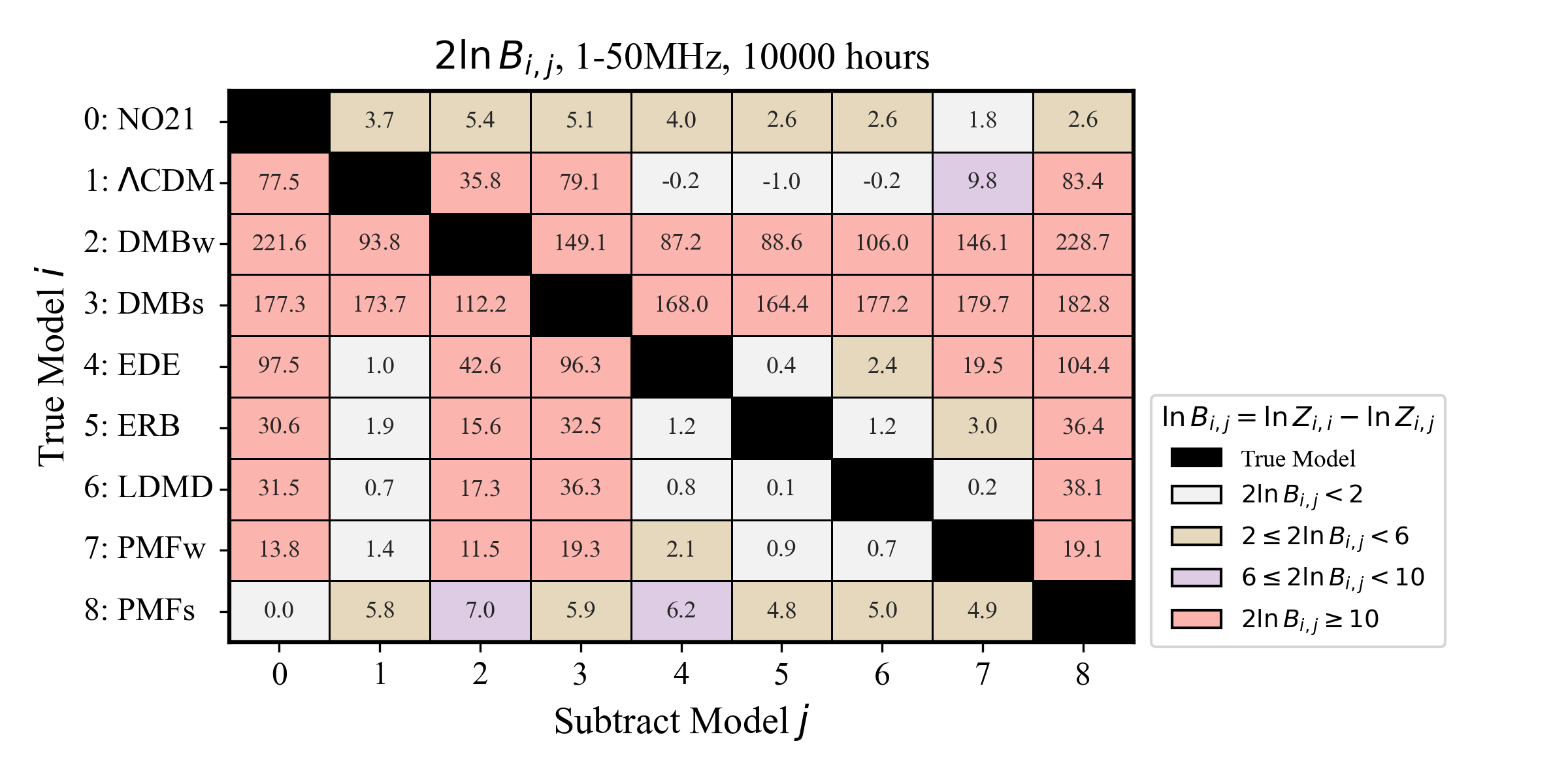}
\caption{$2\ln B_{i,j}$ values for multiple combinations of input and subtraction models assuming a wide-band observation from 1\,MHz to 50\,MHz. Here, $i$ is the input model and $j$ is the subtraction model.}
\label{fig:cont_logZ_Nf50_TOBS10000}
\end{figure*}

Figure~\ref{fig:cont_logZ_Nf50_TOBS10000} summarizes $\ln B_{i,j} = \ln Z_i - \ln Z_j$ for all combinations of true model $i$ and fitted model $j$ for a wide-band observation covering 1\,MHz--50\,MHz with 10,000 h integration time. \syv{For $j=0$, we omit parameter $A$ as same as in previous section.} The mock data contain the 21\,cm signal corresponding to model~$i$. We compare the evidence $\ln Z_i$ obtained when model $i$ is assumed in the fit with the evidence $\ln Z_j$ obtained when model $j$ is assumed in the fit. If $2\ln B_{i,j} > 2$, model $i$ is favored over model $j$. The color of each cell indicates the magnitude of $2\ln B_{i,j}$.

\begin{figure*}
\centering
\begin{flushleft}
\includegraphics[width=8.5cm]{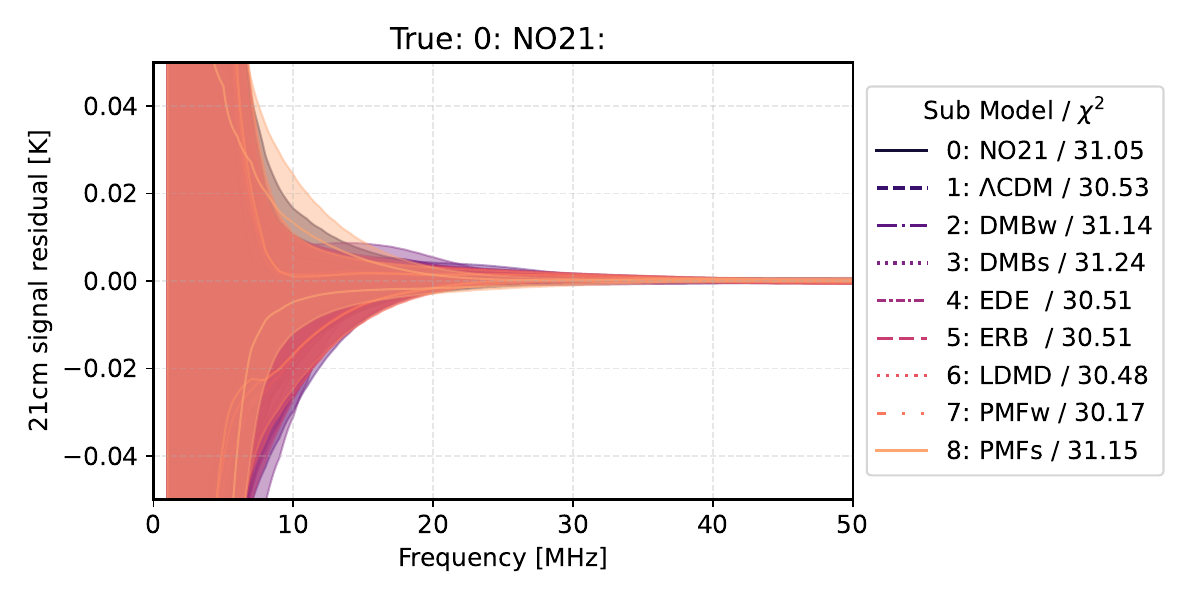}
\includegraphics[width=8.5cm]{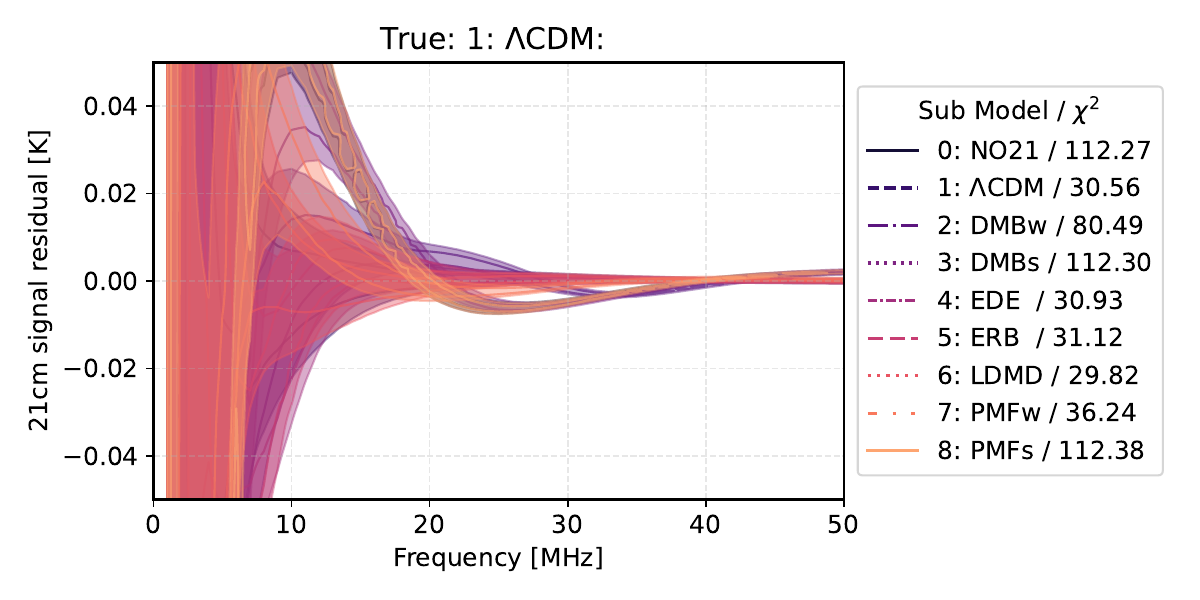}
\includegraphics[width=8.5cm]{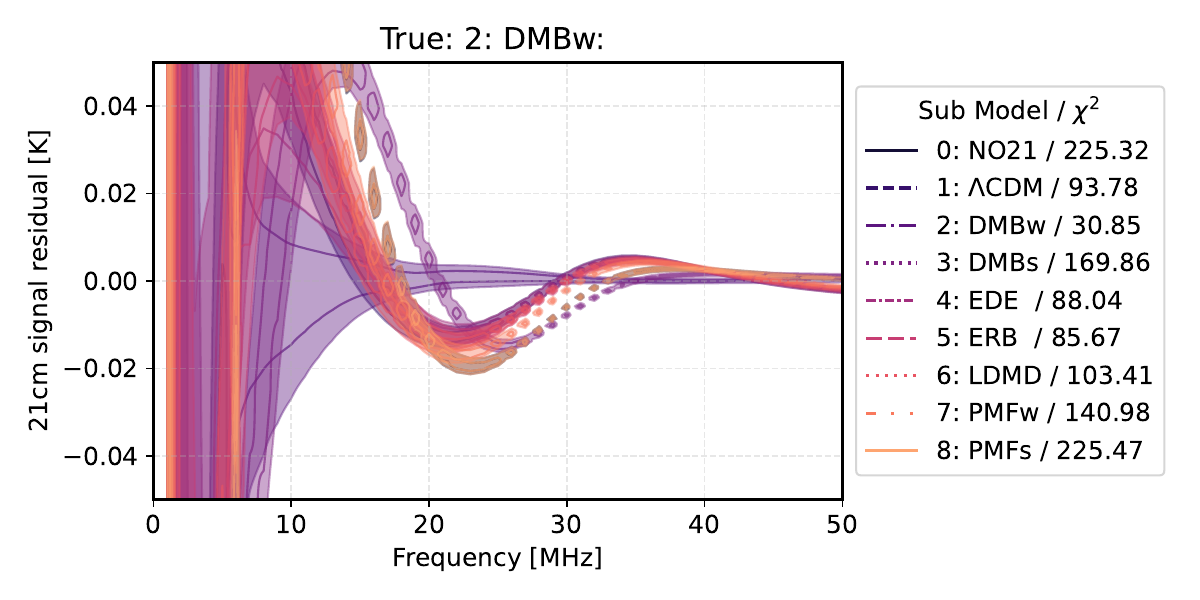}
\includegraphics[width=8.5cm]{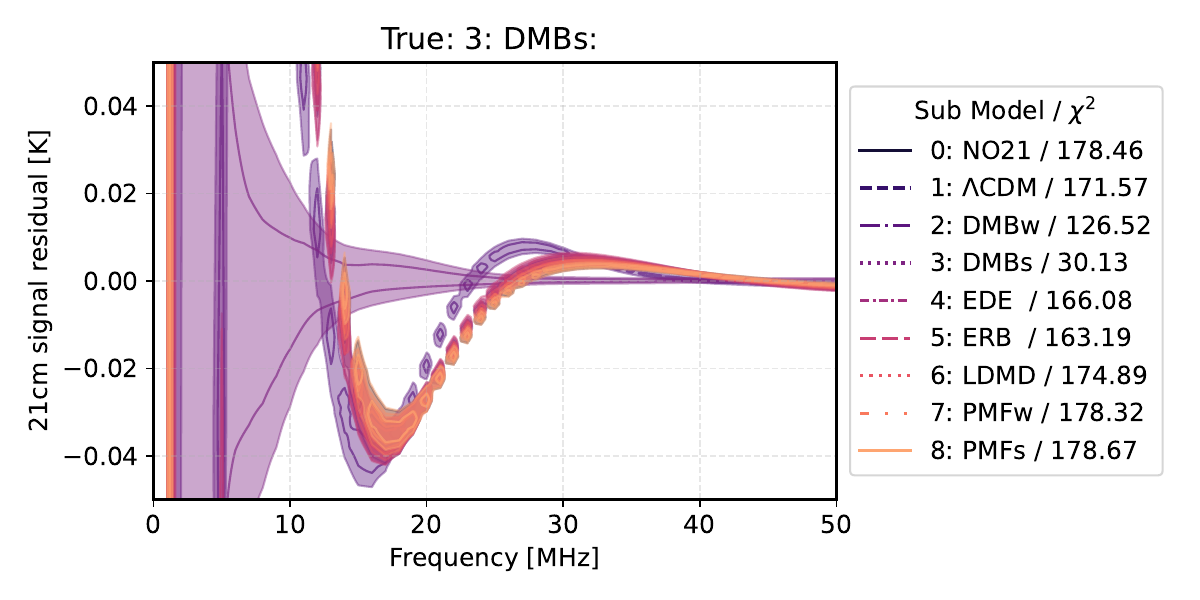}
\includegraphics[width=8.5cm]{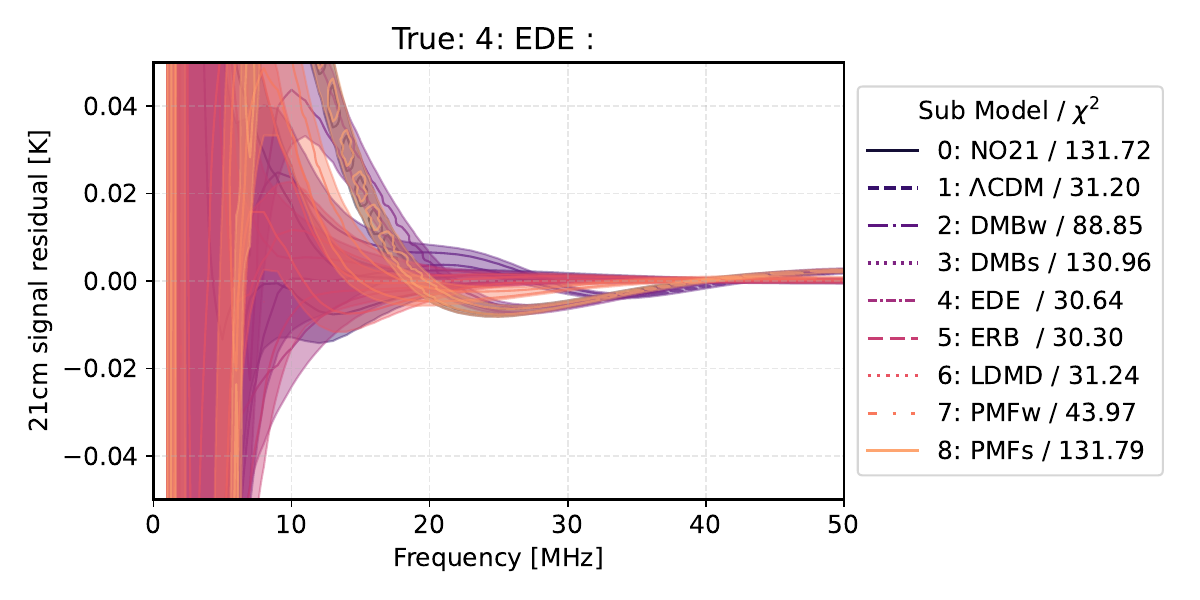}
\includegraphics[width=8.5cm]{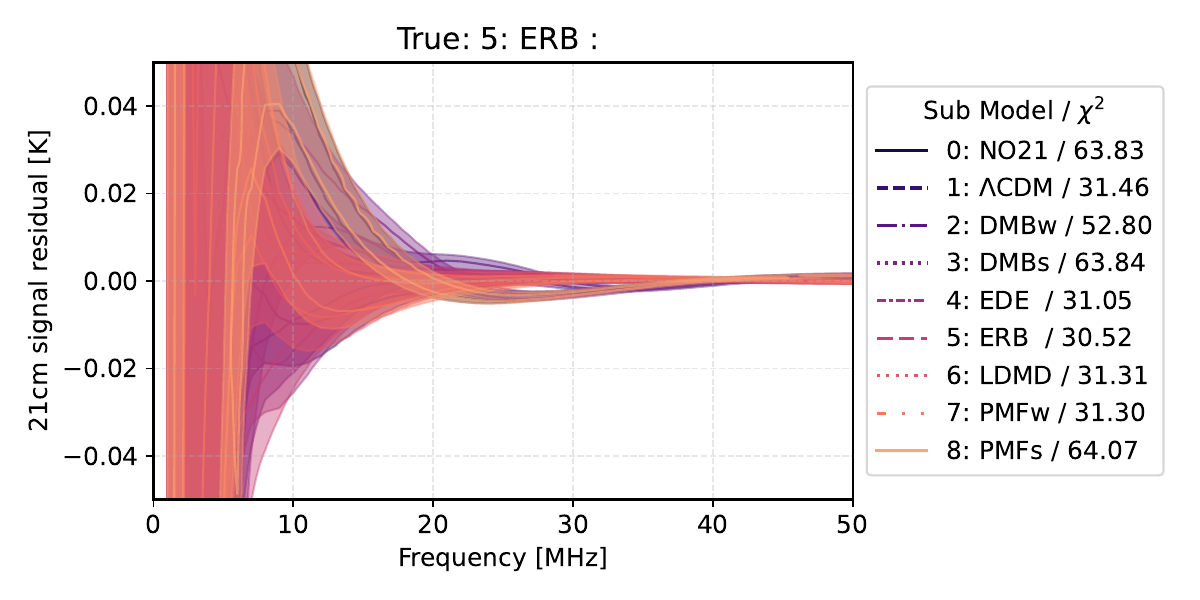}
\includegraphics[width=8.5cm]{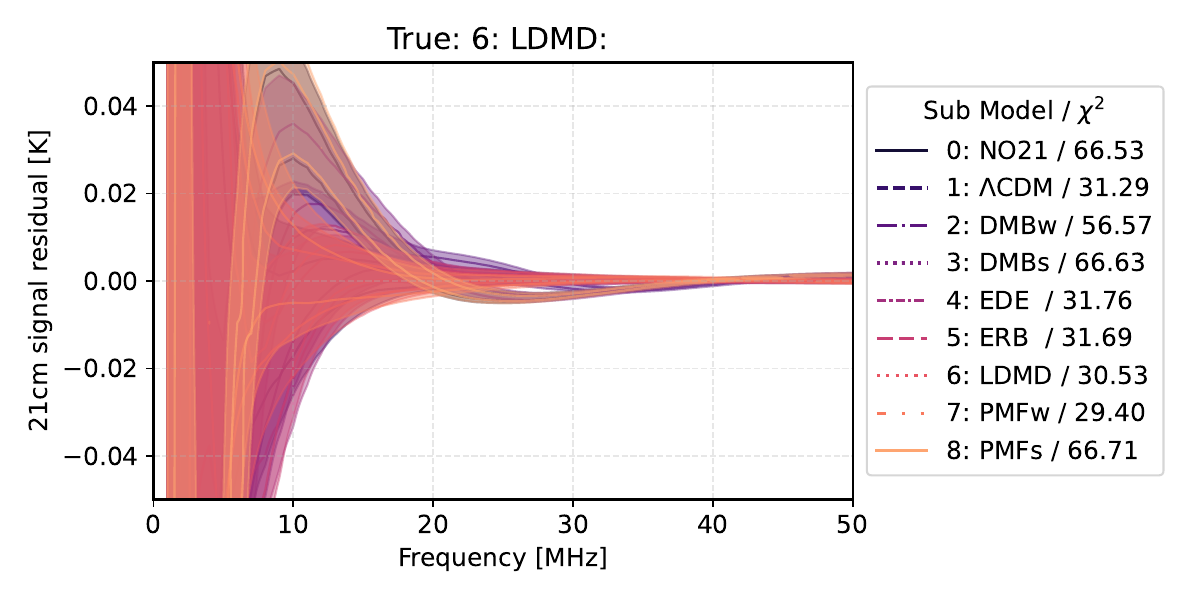}
\includegraphics[width=8.5cm]{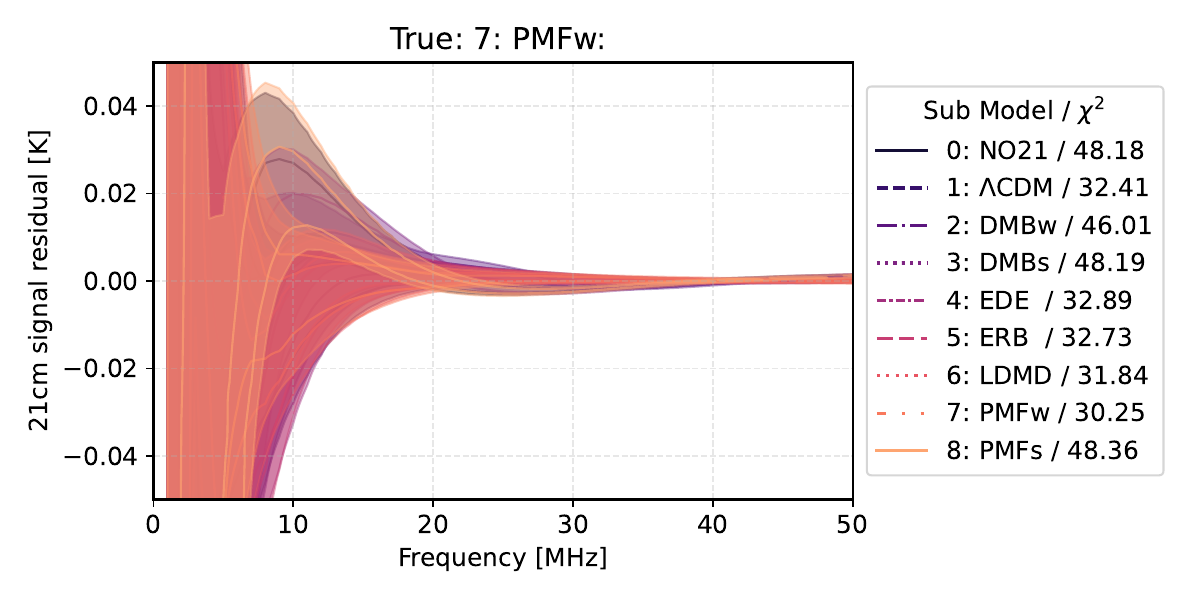}
\includegraphics[width=8.5cm]{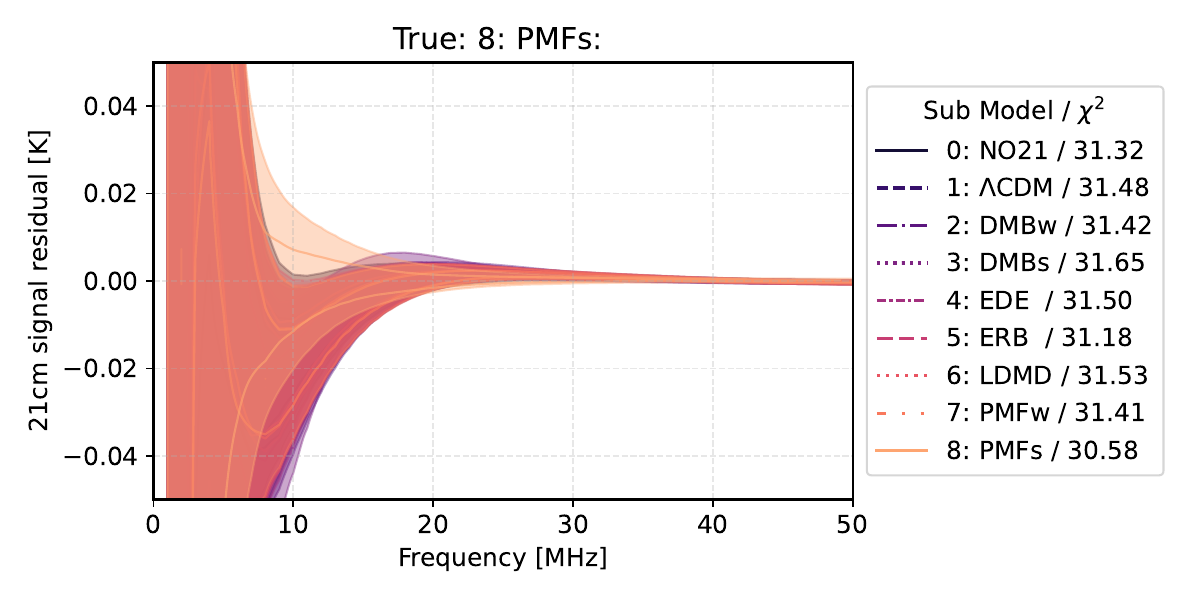}
\end{flushleft}
\caption{Residual after signal and foreground subtraction with maximum likelihood sample. The $\chi^2$ value is also listed in label. Assuming 10,000 hours from 1MHz to 50MHz. The contour shows 1 $\sigma$ error which is plotted using fgivenx \citep{fgivenx}.}
\label{fig:leastsquare}
\end{figure*}

To aid interpretation of the results, we plot the residuals after model subtraction using the maximum-likelihood parameters in Figure~\ref{fig:leastsquare}. In addition, we also calculate the value of $\chi^2$, given as
\begin{equation}
\chi_{i,j}^2 = \sum_k \frac{\left(D_{i,k}-M_{j,k}(\Theta)\right)^2}{\sigma_k^2} \,,
\end{equation}
which is also shown in the legend of the figure. In most cases with $2\ln B_{i,j}>2$, the $\chi_{i,i}^2$ is \syv{well lower} than \syv{$\chi_{i,j}^2$} and the residuals exhibit peaks within 5--30\,MHz. This \syv{reinforces} that $2\ln B_{i,j}>2$ provides a reliable criterion for model discriminability.

Now let us look at each true model in Figure~\ref{fig:leastsquare}. For the NO21 case ($i=0$), the true model is favored over all models except PMFw. \syv{However, the significance of the detection is not strong. As the prior of amplitude parameter $A$ includes $A=0$ and all models can reproduce NO21 case, ideally all models should be indistinguishable. However, $A=0$ lies at the boundary of the prior, and this prior-boundary effect can introduce small difference of $\ln Z$.} As shown in the top-left panel of Figure~\ref{fig:leastsquare}, the residuals are consistent with zero at $\nu > 20$ MHz. 

For the $\Lambda$CDM case ($i=1$), $\Lambda$CDM is not distinguishable from \syv{EDE, ERB and LDMD}. This reflects the similarity of the 21 cm signals in $\Lambda$CDM with \TT{those in} EDE and LDMD, and the residuals are consistent with zero (top-right panel of Figure~\ref{fig:leastsquare}). \syv{For the ERB model, the combination of the foreground parameters, the 21\,cm signal amplitude, and the excess radio background can reproduce the mock spectrum with residuals comparable to those obtained for the true model. Consequently, the posterior distributions of the parameters are mutually consistent for the $\Lambda$CDM and EDE (LDMD) templates, whereas the inferred parameters are biased when the ERB template ($j=5$) is assumed.}

For the DMBw ($i=2$) and DMBs ($i=3$) cases, very strong \syv{template distinguishability} is found against all other models. The residuals exhibit negative peaks around 21 MHz and 18 MHz, and positive peaks around 35 MHz and 29 MHz for DMBw and DMBs, respectively (second row of Figure~\ref{fig:leastsquare}).

For the EDE case ($i=4$), because the 21\,cm signal is similar to $\Lambda$CDM, \syv{the signal cannot be distinguished from either the $\Lambda$CDM or ERB template. In contrast, the LDMD template does not reproduce the EDE signal perfectly, although the statistical significance of the difference remains low.} 

\syv{The ERB ($i=5$) cannot be distinguished from $\Lambda$CDM, EDE and LDMD. Although the ERB signal shape differs from that of the true model, the combination of the foreground parameters, the 21\,cm signal amplitude, and the excess radio background can reproduce the mock spectrum with residuals comparable to those obtained for the true model ($j=i=5$). }

\syv{The LDMD signal ($i=6$) is weak and has a spectral shape similar to that of the PMFw signal. Consequently, the LDMD case cannot be distinguished significantly from the $\Lambda$CDM, EDE, ERB, or PMFw templates.}

\syv{The PMFw signal is also similar in shape to the LDMD signal, but has a smaller amplitude. Consequently, when PMFw is the true model ($i=7$), it cannot be distinguished from the $\Lambda$CDM, ERB, or PMFw templates. Although the significance is low, the PMFw signal can be distinguished from the EDE template. This may be caused by the difference in the redshift of the absorption minimum, which peaked at approximately $z=17$ for PMFw and $z=14$ for EDE.}

For the PMFs case $(i=8)$, because the true 21\,cm signal is positive and peaks at lower frequencies, the observation has sufficient \syv{template distinguishability}. However, \syv{same as $j=0$ case, with $A$=0, the PMFs signal can be absorbed in the foreground parameters. Thus, the significance is low. Clearly,} the signal can be absorbed by the foreground, the residuals are consistent with zero for the $j=0$ (NO21) case. \syv{On the other hand, for $j=8$, the PMFs cannot represent other 21\,cm signals, and therefore the model can be readily disfavored. }

\begin{figure*}
\centering
\includegraphics[width=15cm]{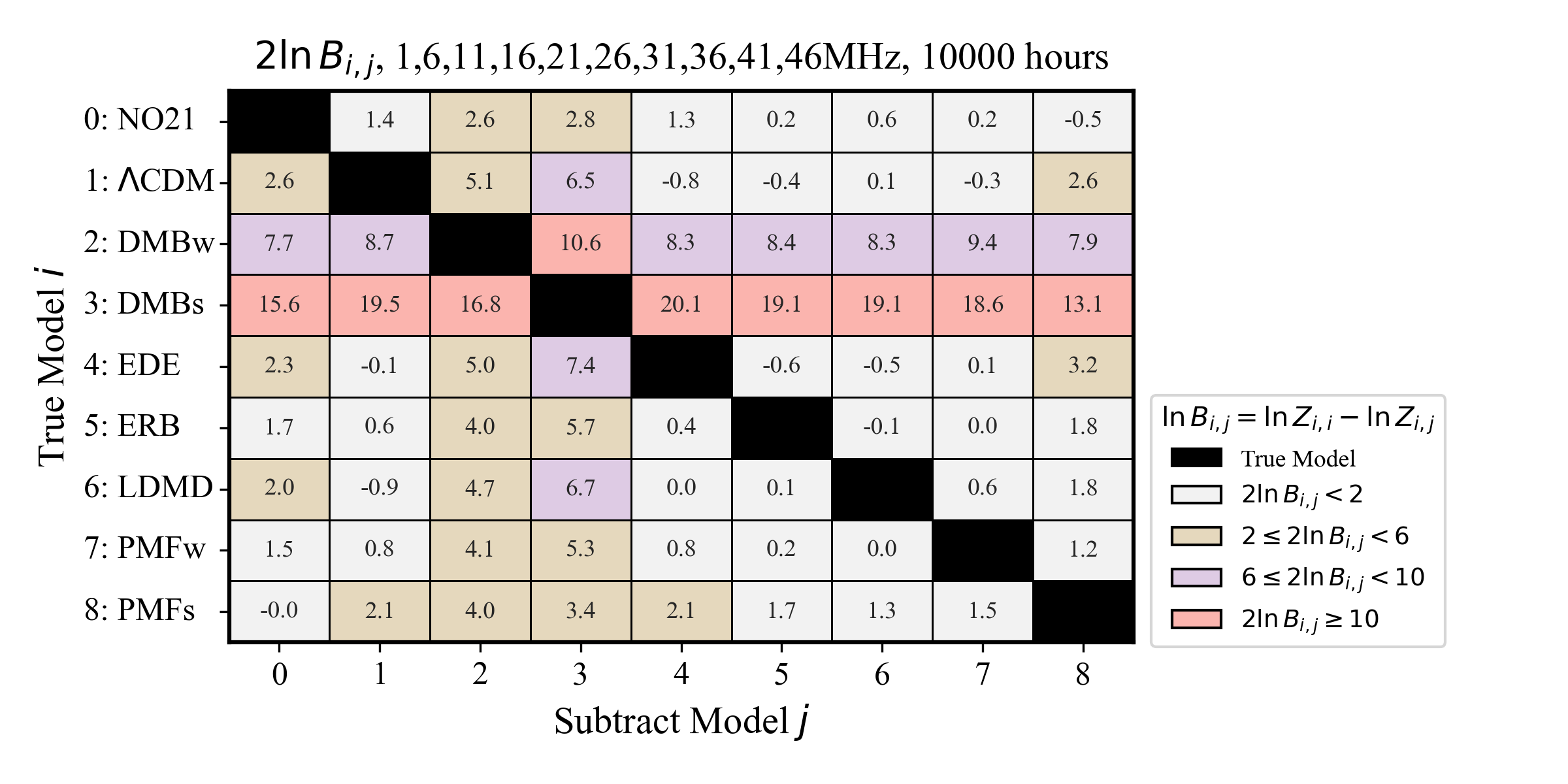}
\caption{Same as Figure~\ref{fig:cont_logZ_Nf50_TOBS10000}, but assuming observation at 1, 5, 10, 15, 20, 25, 30, 35, 40, 45, 50 MHz. }
\label{fig:cont_logZ_Nf11_TOBS10000}
\end{figure*}

We next consider observations at 5\,MHz intervals from 1\,MHz to 46\,MHz.  Figure~\ref{fig:cont_logZ_Nf11_TOBS10000} shows the results. Compared to the wide-band observation, \syv{the template distinguishability} is reduced. In particular, models with weak signals ($\Lambda$CDM, EDE, LDMD, PMFw) and smooth spectra \syv{(ERB and PMFs)} cannot be distinguished under this observing configuration \syv{although there are a few combination which can achieve $2\ln B_{i,j}>2$.} On the otherhand, even with the limited channels, the DMBw and DMBs models remain strongly distinguishable from the all other models.

\begin{figure*}
\centering
\includegraphics[width=15cm]{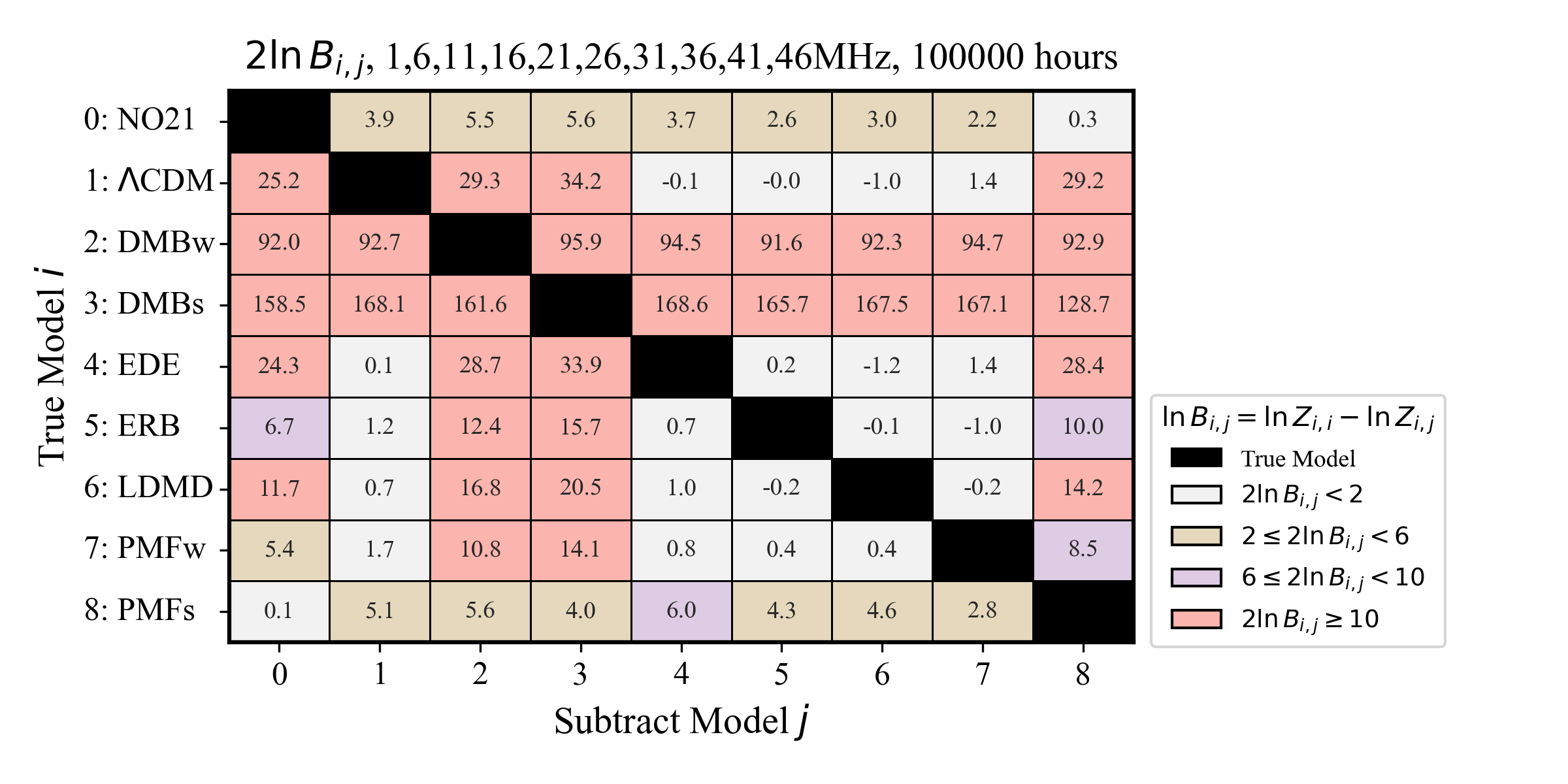}
\caption{Same as Figure~\ref{fig:cont_logZ_Nf11_TOBS10000}, but assuming 100,000 hours of integration time.}
\label{fig:result3}
\end{figure*}

As in the previous subsection, we also analyze an idealized case assuming 100,000 hours of integration time for observations performed at 5\,MHz intervals from 1 MHz to 46 MHz. Figure~\ref{fig:result3} shows the results, which indicate that \TT{the tendency of the template distinguishability is broadly the same as that shown in Figure~\ref{fig:cont_logZ_Nf11_TOBS10000}, but with a larger $2 \ln B_{i,j}$.} We again emphasize that the spectral structure of the 21\,cm signal can be captured with a limited number of frequency channels and distinguished from the foreground. 

To further improve \syv{template distinguishability}, additional information on the foreground would be highly beneficial. For example, SKA1-Low covers frequencies down to 50\,MHz and offers significantly better sensitivity than existing instruments. Such observations could constrain the parameters of the extragalactic foreground temperature at 50\,MHz. Since galactic diffuse emission exhibits an anisotropic distribution, whereas the 21 cm signal is expected to be spatially uniform, the direction-dependent component of the foreground may help foreground removal \cite{2020ApJ...897..175T, 2023MNRAS.520..850A}. In practice, free–free absorption by H\,{\sc ii} regions introduces anisotropic structures in the global 21\,cm spectrum \cite{2022MNRAS.513.5125S}. Therefore, accurate foreground removal requires proper modeling of the sky and error characterization \cite{2021ApJ...914..128C, 2026ApJ...997....4L}. We leave these studies for future work.

In this study, we do not include realistic systematic errors, which may be spectrally non-smooth, because the primary purpose of this work is to assess the detection performance and \syv{template distinguishability} under ideal conditions. In practice, there are several instrumental (e.g., beam effects and calibration) and observational (e.g., ionospheric effects and RFI) systematics. In such cases, wide-band observations can be limited by these systematics, and sophisticated mitigation techniques will be required.

\section{Summary}\label{sec:4}

We assessed the detectability of the 21\,cm signal under several observing strategies. By employing nine 21\,cm models and a physically motivated foreground model, we performed a Bayesian analysis using a nested-sampling algorithm. 
The likelihood was evaluated by varying the foreground parameters \syv{and amplitude of} the 21\,cm signal with \syv{fixed shape of template models} assumed in the fit, which is subtracted from the mock observed data. Based on the Bayes factor analysis, \syv{for comparison}, we \syv{confirmed} that a wide-band observation covering 1--50\,MHz provides strong detection and template distinguishability for most 21\,cm models. Although observations with a limited number of channels suffer from reduced detectability due to insufficient sensitivity, such observations can achieve detection for most models if higher sensitivity is assumed. This implies that a small number of channels can be sufficient to distinguish the 21\,cm signal from the foreground owing to intrinsic differences in spectral shape. 

\syv{Using several observing-frequency strategies and Bayesian-evidence-based model comparison, we confirm the importance of wide-band observations, as emphasized in previous studies \citep[e.g.][]{2010PhRvD..82b3006P,PhysRevD.87.043002,2015ApJ...813...11M,2016MNRAS.455.3829H}. In particular, observations that do not extend below 3\,MHz fail to detect the 21\,cm signal, except for models with very strong absorption features due to degeneracy with the foreground parameters. This result highlights the importance of very-low-frequency coverage for \TT{separating} the free--free absorption component. We also examine the effect of a frequency-independent noise floor on detectability and find that residual errors must be controlled to approximately 10\,mK or below for the $\Lambda$CDM model to be \TT{detected with strong evidence.}
}

\syv{Regarding template discrimination, models with strong absorption features, such as DMBw and DMBs, can generally be distinguished from the other templates. By contrast, signals with similar spectral shapes, including $\Lambda$CDM, EDE, LDMD, and PMFw, are difficult to distinguish from one another, with a few exceptions. The ERB template can reproduce a $\Lambda$CDM-like absorption feature through a combination of the foreground components, the 21\,cm signal, and the excess radio background. Consequently, under our assumptions, the $\Lambda$CDM signal cannot be distinguished from the ERB model. Conversely, the smooth, positive PMFs spectrum is disfavored when the true signal contains an absorption feature that cannot be reproduced by the PMFs template.}

\begin{acknowledgments}
We thank to members of TREED project for their support. We also thank to Teppei Minoda for their fruitful discussion on the modelling of the primoridal magnetic fields. 

This research was supported by the grant of OML Project by the National Institutes of Natural Sciences (NINS program No, OML022303, OML022402, OML022502). This work is partially supported by JSPS KAKENHI (Grant Number 24K17098(SY), 24K00625(SY), 25K01004 (TT)) and MEXT KAKENHI (Grant Number 23H04515 (TT), 25H01543 (TT)). F.O.~is supported by individual research funding from Nihon University.

We acknowledge the use of an AI-based writing tool for editing and coding. 

\end{acknowledgments}

\bibliography{bibtex_this}

\end{document}